\documentclass[journal,article,submit,pdftex,moreauthors]{Definitions/mdpi} 
\nolinenumbers
\firstpage{1} 
\pubvolume{1}
\issuenum{1}
\articlenumber{0}
\pubyear{2025}
\copyrightyear{2025}
\datereceived{ } 
\daterevised{ } 
\dateaccepted{ } 
\datepublished{ } 
\hreflink{https://doi.org/} 

\def \arcmin      {\text{$^\prime$}}
\def \arcsec      {\text{$^{\prime\prime}$}}

\Title{RAD@home citizen science discovery of two spiral galaxies where the 30-220 kpc radio lobes are possibly shaped by ram pressure stripping}

\TitleCitation{RAD@home discovery of two ram pressure-stripped spiral-host radio galaxies}

\Author{Prakash Apoorva $^{1,2}$, Ananda Hota $^{3,2}$*, Pratik Dabhade $^{4,2}$, 
P.K. Navaneeth$^{2}$, Dhruv Nayak$^{2}$, Arundhati Purohit$^{2}$}

\AuthorNames{Firstname Lastname, Firstname Lastname and Firstname Lastname}

\address{%
$^{1}$ \quad Department of Physics, Faculty of Engineering and Technology, Jain Deemed-to-be University, Kanakapura Road, Bangalore 562 112, Karnataka, India\\
$^{2}$ \quad RAD@home Astronomy Collaboratory, Kharghar, Navi Mumbai, India; hotaananda@gmail.com \\
$^{3}$ \quad UM-DAE Centre for Excellence in Basic Sciences, University of Mumbai, Kalina, Mumbai, India\\
$^{4}$ \quad Astrophysics division, National Centre for Nuclear Research, Pasteura 7, 02-093 Warsaw, Poland}

\corres{Correspondence: hotaananda@gmail.com
}

\abstract{We report the RAD@home citizen science discovery of two rare spiral-host radio galaxies (NGC 3898 \& WISEA J221656.57-132042434.1 or RAD-``Thumbs up'' galaxy), both exhibiting asymmetric radio lobes extending over 30 to 220 kiloparsec scales. We present a multi-wavelength image analysis of these two sources using radio, optical, and ultraviolet data. Both host galaxies are young, star-forming systems with asymmetric or distorted stellar disks. These disks show similarities to those in galaxies undergoing ram pressure stripping, and the radio morphologies resemble those of asymmetric or bent FR-II and wide-angle-tailed radio galaxies. We suggest that non-uniform gas density in the environment surrounding the ram pressure-stripped disks may contribute to the observed asymmetry in the size, shape, and brightness of the bipolar radio lobes. Such environmental effects, when properly accounted for, could help explain many of the non-standard radio morphologies observed in Seyfert galaxies and in recently identified populations of galaxies with galaxy-scale radio jets, now being revealed through deep and sensitive radio surveys with uGMRT, MeerKAT, LOFAR, and, in the future, the SKAO. These findings also underscore the potential of citizen science to complement professional research and data-driven approaches involving machine learning and artificial intelligence in the analysis of complex radio sources.}


\begin{document}



\nolinenumbers
\section{Introduction} 

Double-lobed radio galaxies were first discovered in 1953, and since then, they have been almost exclusively linked to elliptical hosts. Only in the past $\sim$\,15~years has a modest but growing sample demonstrated that spiral and disk galaxies can also launch episodic, megaparsec-scale radio jets \citep{Hota2011,Bagchi2014,Sethi2025}. While many nearby Seyferts in spiral galaxies are known to host radio jets and lobes, although a few kpc-scale, \citep[]{Baum, Colbert, Hota2006, Gallimore}, only recently, due to the advancement of sensitive low-frequency surveys, a sample of galaxy-scale radio-lobed galaxies has been discovered \citep{Webster2021}. Hence, for a unified picture of radio jets in kpc-to-Mpc scale and in bulge-less spirals to giant ellipticals, the study of jets at intermediate scales are important. Since these galaxy-scale jets in spiral hosts are naturally easier to observe for jet-interstellar Medium (jet-ISM) and jet-circum-galactic Medium (jet-CGM) interactions, their implications for understanding detailed physics of AGN-feedback in galaxy evolution are naturally high. Large, 200 kpc to Mpc-scale radio lobes are likely too large to show signs of interaction with the CGM. Small kpc-scale Seyfert jets are easy to show interacting with the ISM. In some cases of jet-ISM interactions, jet-driven shock ionisation is clearly observed \citep[]{Cecil, Capetti}. As in the well-known case of Minkowski's Object, jet-triggered young star formations have also been observed in many galaxies. Radio jets are always seen orthogonal to the small-scale or nuclear dust lanes in elliptical galaxies, irrespective of cases where the large-scale dust ellipse is aligned with the major axis of the optical galaxy \citep{VerdoesKleijn}. Such orthogonal correlations naturally suggest gas/dust feeding of the accretion disks. However, jets in spiral galaxies do not show any such correlation with the rotation axis \citep{Kinney}. This can be attributed to intense jet-ISM interaction and short duty cycles of accretion and jet episodes. In some cases, instead of well-defined jets, radio emissions from spirals are seen in the shape of bipolar bubbles. In cases where bubbles are seen to be orthogonal to the stellar disk, bubbles are proposed to have been shaped by the galactic wind outflow \citep{Baum}. In a representative sample of ten radio bubbles, the mandatory presence of AGNs has been observed \citep{Hota2006}. The radio plasma in the bubble has probably been supplied by the AGN jet. The current study reports the finding of two galaxies where the intermediate scale (30-200 kpc) radio lobes from two spiral galaxies are seen asymmetrically located w.r.t. the host galaxies. Taking clues from the multi-wavelength observations, we bring an analogy with the ram pressure stripped galaxies \citep[]{GunnStripping, AbadiStripping} and propose a novel idea that these intermediate-scale radio lobes seen here, and possibly in other cases, can be explained by the incorporation of ram pressure stripping by the intra-group or intra-cluster medium. Discovery and study of more such cases, added with deep H\,{\sc i} and CO observations, will grow our understanding of radio-jet feedback-driven galaxy evolution \citep[]{Croton, Hardcastle}. \\

\section{RAD@home citizen science discovery method}
 
The discovery and interpretation of the sources presented in this study were carried out through the RAD@home Citizen Science Research (CSR) Collaboratory \citep[]{Hota2016, Hota2022}. Of over 4700 members, thousands were trained to interpret galaxy images through both online and in-person interactions. Undergraduate students across India were given the opportunity to explore and understand the fundamentals of multi-wavelength astronomy using Red-Green-Blue-Contour (RGB-C) images constructed from ultraviolet (UV), optical, infrared (IR), and radio data. To introduce the basic concepts of star formation, IR–optical–UV (IOU) RGB images were used, incorporating IR data from the Wide-field Infrared Survey Explorer \citep[WISE;][]{WISE}, UV data from the Galaxy Evolution Explorer \citep[GALEX;][]{GALEX}, and optical images from the Digitised Sky Survey (DSS). Participants were encouraged to analyse RGB-C images of different galaxies and share their interpretations on various days throughout the year. These analyses were further discussed through text and image comments, as well as interactive online sessions. Regular e-classes were conducted via Google Meet on weekends, providing a platform for collaborative learning and feedback. A custom multi-wavelength RGB-C image maker tool (see Fig.~\ref{fig:ngc3898} for an example) ensured consistent and foundational training in galaxy image interpretation \citep{Kumar2021}.

Upon successfully completing the initial RGB-C image analysis training, students were further guided to work with FITS image files from the TIFR GMRT Sky Survey (TGSS), conducted with the Giant Metrewave Radio Telescope (GMRT) at 150~MHz, offering a resolution of $\sim$\,25$^{\prime\prime}$ and an rms noise of $\sim$\,5 mJy beam$^{-1}$ \citep{TGSS}. The primary goal of this exercise, performed using SAO~ds9, was to identify faint, diffuse, and non-standard radio structures.
Candidate sources identified by trained citizen scientists were discussed collaboratively through radio–optical–radio overlay images generated using the RAD@home RGB-maker tool. To protect the interests of the discoverer, the coordinates of the sources were withheld during the discussion. This process also helped ensure consistent multi-wavelength image analysis skills development for all participants. The faint, diffuse structures may represent remnants of past AGN jet activity \citep{Hota2016}. Non-standard morphologies, on the other hand, could indicate unusual AGN interactions or environmental distortions, such as in the case of RAD12 \citep{Hota2022}, where a radio jet from one galaxy impacts a companion galaxy and rebounds to form a mushroom-shaped radio bubble. Discoveries like these offer valuable insights into galaxy evolution, star formation, and AGN feedback via jet-driven outflows.
Promising candidates identified through this process are submitted to the Collaboratory via a dedicated image submission Google form, after which they are reviewed by professional astronomers and selected for follow-up observations with GMRT or direct publication. In this study, we report on two such galaxies exhibiting non-standard and faint, diffuse radio structures.


\section{Results}

\subsection{Radio lobes of NGC3898 or a background radio galaxy ?}
\begin{figure}[H]
\includegraphics[width=12 cm]{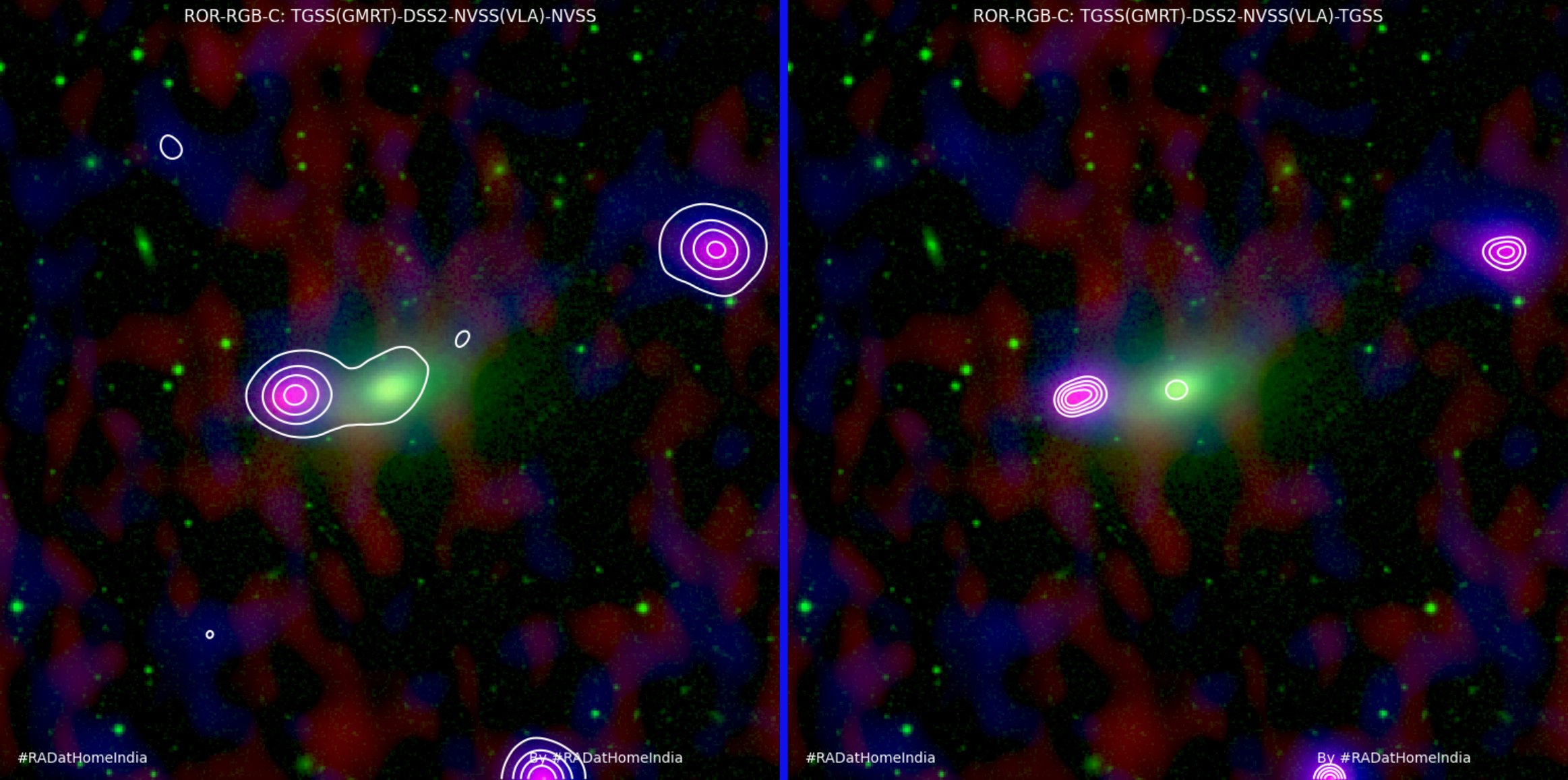}
\caption{ A typical RAD@home citizen science research RGB-maker web-tool output showing radio-optical-radio overlay of the galaxy NGC3898. The Red, Green, and Blue channels present TGSS-ADR1 (150 MHz), DSS-R and NVSS (1400 MHz) images, respectively. While the left panel shows NVSS in contour [0.0015, 0.0045, 0.0076, 0.0106 Jy beam$^{-1}$ ] the right panel shows TGSS contours [0.015, 0.023, 0.03, 0.038 Jy beam$^{-1}$]. In both cases, the contours start with 3 times the r.m.s. noise in each image. While NVSS beam is 45$^{\prime\prime}$ the TGSS beam is 25$^{\prime\prime}$. \label{fig:ngc3898}}
\end{figure}   

Two radio blobs were seen in the TGSS image, but the possible host galaxy, a spiral galaxy, was not in the middle, but close to one of the blobs. This extreme case was posted on 21 April 2024 in the RAD@home online discussion group. The RGB-contour image constructed using radio-optical-radio data (Fig.~\ref{fig:ngc3898}) shows the eastern lobe to be located close to the possible host galaxy, NGC3898, while the western lobe is nearly three times farther away. Although the shorter eastern lobe is relatively bright, the western lobe appears faint and disconnected from the possible host. Optical disk of NGC3898 also exhibits faint radio emission in the NRAO VLA Sky Survey (NVSS; 1400~MHz, beam size = 45$^{\prime\prime}$) data \citep{NVSS}. A comparison with the Faint Images of the Radio Sky at Twenty Centimeters (FIRST; 1400~MHz, beam size = 5$^{\prime\prime}$) survey \citep{FIRST} reveals a compact radio source coincident with the nucleus of NGC~3898 ($z_{spec} = 0.003875$, $v = 1162 \pm 1$~km\,s$^{-1}$, scale = 0.081~kpc\,arcsec$^{-1}$ or 1$^{\prime}$=4.86 kpc). We obtained the spectral index values  ($\alpha_{150~\mathrm{MHz}}^{1400~\mathrm{MHz}}$) for the target from the SPIDX database \citep{SPIDX}, which provides spectral index derived from the NVSS and reprocessed TGSS data. In this dataset, UV-tapering was applied to the original TGSS visibilities (original beam $\sim$25$^{\prime\prime}$) to match the coarser resolution of the NVSS ($\sim$45$^{\prime\prime}$), ensuring consistent spatial sampling across the two frequencies. The eastern lobe, central component and western lobes have average spectral indices of -0.79, -1.0 and -0.58, respectively. 

NGC~3898 (UGC~06787) is a well-studied nearby galaxy, with observations spanning nearly all wavelengths. This SA(s)ab-type spiral galaxy, classified as a LINER--H\,{\sc ii} composite, is a member of the ``12-3'' galaxy group \citep{NGC3898group}. NGC~3898 is also identified as a galaxy (ID 461333) in the catalogue of merging groups and clusters \citep{Tempel2017}. It belongs to Group ID 52914, which comprises 30 member galaxies, with an $R_{200}$ of 580~kpc and a total mass $M_{200} = 2.1 \times 10^{13}\ M_\odot$. Likely due to the presence of extremely asymmetric radio emission blobs seen in NVSS, the galaxy has been overlooked in earlier samples of nearby galaxies hosting radio jets. In this brief study, we focus on the asymmetric radio lobe morphology and complement the analysis with newly available data.

\begin{figure}[H]
\includegraphics[width=12 cm]{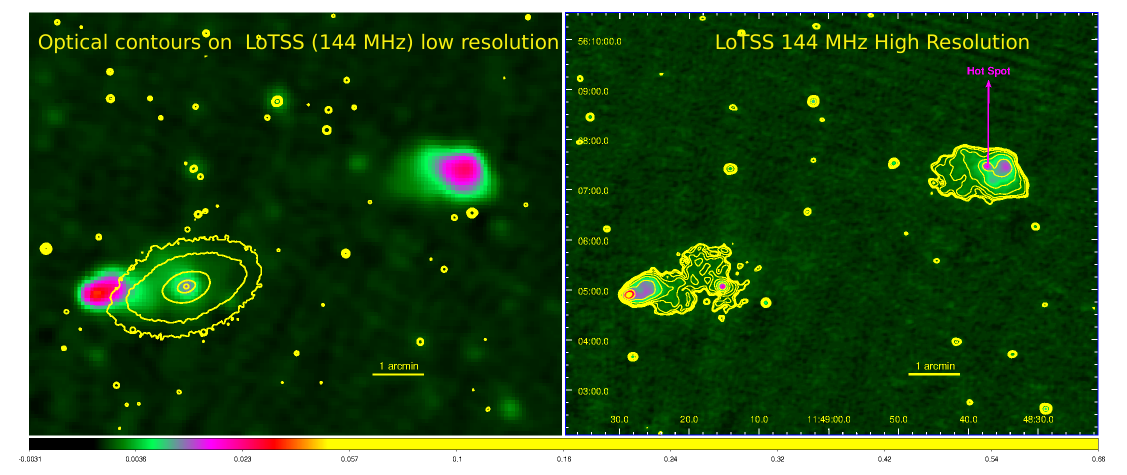}
\caption{ NGC3898: Optical image (SDSS r-band) in thick yellow contour is plotted on the LoTSS (144 MHz beam=20$^{\prime\prime}$) low-resolution radio image from LoTSS (left panel).  The LoTSS high resolution (beam=6$^{\prime\prime}$) false colour image (right panel) is shown along with a contour (yellow thin). The contour levels, in log scale, are 0.00024, 0.00025, 0.00027, 0.00033, 0.00044, 0.00068, 0.0012, 0.0023, 0.0047, 0.0099 Jy beam$^{-1}$. The contours start with 3 times the r.m.s. noise in the radio image. The scale bar for 1$^{\prime}$=4.86 kpc has been put at the right-hand bottom of the image. The point radio source (R.A: 11:49:03.105 Dec. +56:06:33.049) located in the geometric mid-point of the radio lobes has been presented, in detail, in the next Figure. \label{fig2}}
\end{figure}   

Radio images are also available for this target from the LOFAR Two-metre Sky Survey \citep[LoTSS;][]{LoTSS}. In Fig.~\ref{fig2}, we present both low (20\arcsec) and high-resolution (6\arcsec) 144~MHz images from LoTSS. In the high resolution image (right panel) the eastern lobe reveals a clearly resolved structure, with a hotspot and a backflow forming a standard edge-brightened FR-II morphology resembling a ``fish-head'' or bow shock-like structure. The hotspot is located at the farthest end from NGC3898. The projected angular separation between the hotspot and the radio core, nucleus of NGC3898, is 113$^{\prime\prime}$, corresponding to a projected size of 9~kpc assuming the radio lobes are associated with NGC3898. Faint radio emission from the backflow appears to reach NGC3898. In contrast, the western lobe is well separated from the spiral galaxy and appears distorted. The western hotspot is offset to the north relative to the remaining diffuse emission of the lobe. Extended radio emission is observed farther from the spiral galaxy than the hotspot. The projected distance between the western hotspot and the radio core of the spiral is 349$^{\prime\prime}$ or, assuming the radio lobe and spiral association, 28~kpc. The resulting arm-length ratio, greater than 3, makes this an extreme example of lobe asymmetry. The total flux density of the radio core region is 10~mJy, with a peak flux of 5.73~mJy\,beam$^{-1}$. The eastern lobe, including the radio core of the spiral, has a flux density of 175~mJy, while the western lobe has 141~mJy. Since the eastern lobe extends into the spiral galaxy, the core flux must be subtracted when estimating the true flux density of the lobe. It remains uncertain whether ongoing star formation in the galactic disc contributes to the measured flux density of the eastern lobe.

The low-resolution LoTSS image is shown in Fig.~\ref{fig2} (left panel), with optical image contours overlaid. No additional features are evident, except that the eastern lobe's backflow is clearly seen crossing the nucleus of the spiral galaxy. The backflow also extends twice as far to the north of the optical nucleus compared to the south. This suggests that the eastern backflow is unlikely to be interacting directly with the spiral galaxy but appears to do so primarily due to projection effects. The origin of this pronounced asymmetry will be discussed in a subsequent section.

\begin{figure}[H]
\includegraphics[width=7cm]{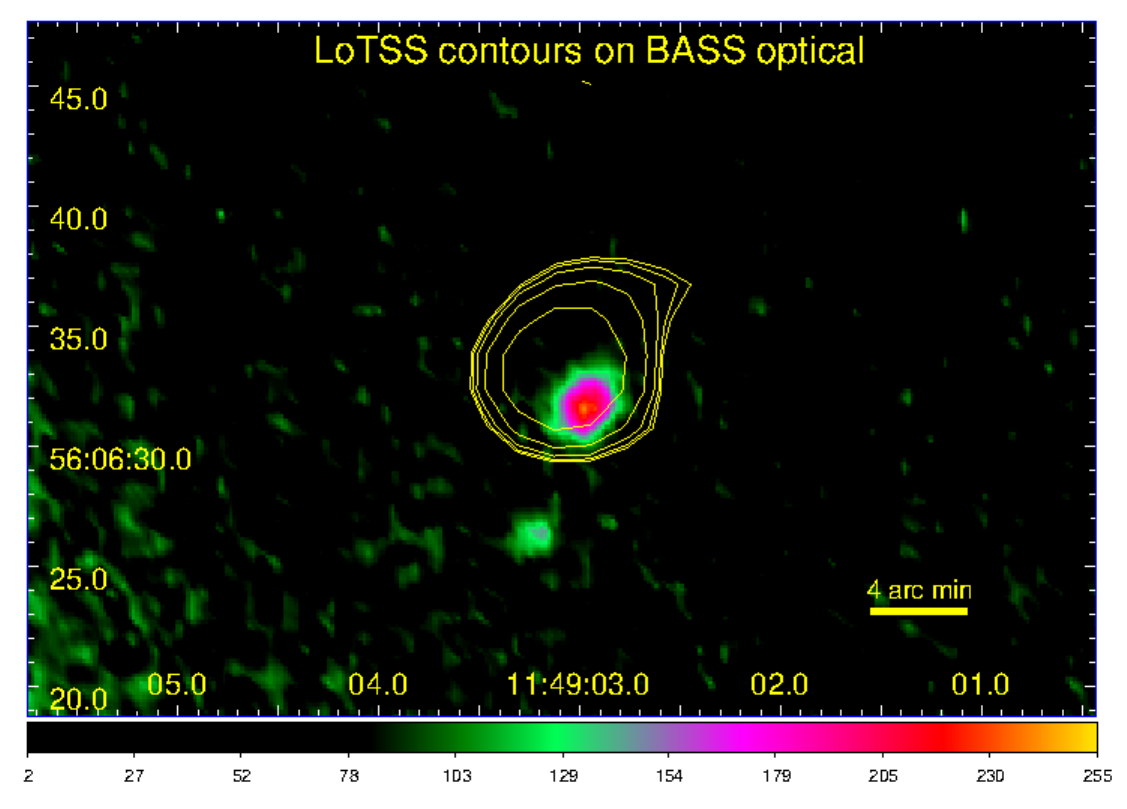}
\includegraphics[width=5.5cm]{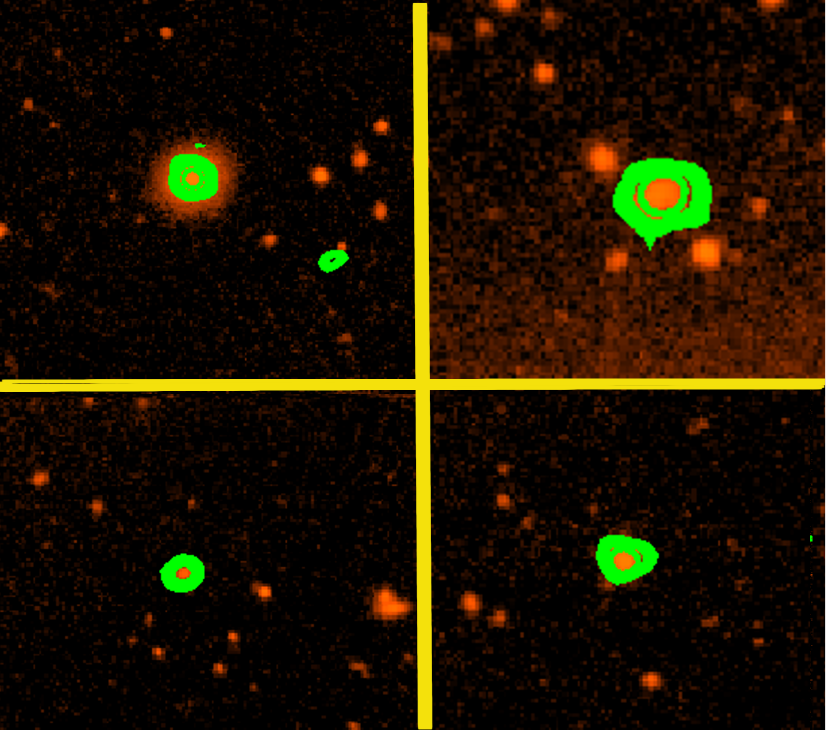}
\caption{Left panel: LoTSS 6\arcsec angular resolution image contours (same as in previous figure) are plotted over optical image from the BASS survey. The LoTSS point source (R.A: 11:49:03.105 Dec. +56:06:33.049) is the only radio source seen between the radio lobes. The scale bar for 4$^{\prime\prime}$ has been put at the right-hand bottom of the image. Over 2$^{\prime\prime}$ positional offset is a clear suggestion that it is not the radio core of the radio lobes discussed here, as associated with NGC3898. Right panel: The collage presents four point sources in radio and optical seen in four directions around NGC3898. High resolution LoTSS contours (green) are plotted on BASS optical image to demonstrate the astrometric accuracy of the images. \label{fig:NGC3898-LOFAR-Legacy-not-host-ds9}
}
\end{figure}   

To investigate further if the asymmetric radio lobes are indeed associated with  NGC~3898 and it's not a background radio galaxy, we examined the LoTSS 6\arcsec-resolution radio image alongside the optical data from the Beijing-Arizona Sky Survey \citep[BASS;][]{BASS}. Near the midpoint between the two radio lobes, we identify a faint, compact radio source (R.A. 11:49:03.105, Dec. +56:06:33.049) that could potentially be the radio core of the double lobed radio source. Upon closer inspection, we find that the peak of this compact radio source is offset by more than 2\arcsec from the nearest optical galaxy (R.A. 11:49:02.958, Dec. +56:06:31.398), which has a photometric redshift of $z_{phot} = 0.469 \pm 0.091$ (see Fig.~\ref{fig:NGC3898-LOFAR-Legacy-not-host-ds9} left panel). The positional accuracy of LoTSS 6\arcsec-resolution radio images are within 0.2\arcsec \citep{LOFARaccuracyShimwell}. At $z =$ 0.469, the 2\arcsec separation corresponds to nearly 12 kpc between the radio point source and the optical galaxy, assuming that the radio source is at the distance of the galaxy. Furthermore, we also verified the astrometric accuracy of the LoTSS image by cross-matching point sources across the field. The right panel of the same figure shows a collage of four such point radio sources whose LoTSS contours (green) are overlaid on the BASS optical image. The compact radio source (left panel) has a flux density of 0.8~mJy at 144~MHz but is not detected in higher frequency radio surveys, including FIRST (3$\sigma$ upper limit: 0.5~mJy) and Very Large Array Sky Survey(VLASS; 3$\sigma$ upper limit: 0.4~mJy). This leads to an upper limit in the spectral index $\alpha_{150~\mathrm{MHz}}^{1400~\mathrm{MHz}} <$ -0.2. This value is consistent with it being a flat-spectrum radio core as well as with a steep-spectrum radio galaxy at an even higher redshift whose optical host is undetected in BASS. If the double lobe structure is a background radio galaxy with its host galaxy at $z_{phot} = 0.469 \pm 0.091$, the total size (113$^{\prime\prime}$+349$^{\prime\prime}$=462$^{\prime\prime}$; scale=5.95 kpc/$^{\prime\prime}$) is $\sim$\,2.75 Mpc. Radio galaxies with sizes greater than 700 kpc are called giant radio galaxies (GRGs), which are a relatively rare population \citep{GRSREV}. In summary, the association of the LoTSS point source with the double radio lobes is not proven due to the lack of detection of a radio jet aligned with the lobes, and the compact radio source being significantly offset from the centre of the nearest optical galaxy. Based on this analysis, we interpret that the asymmetric radio lobes are associated with NGC~3898 (hereafter the host galaxy) rather than with a background galaxy.
 

\begin{figure}[H]
\begin{adjustwidth}{-\extralength}{0cm}
\centering
\subfloat[\centering]{\includegraphics[width=6.0cm]{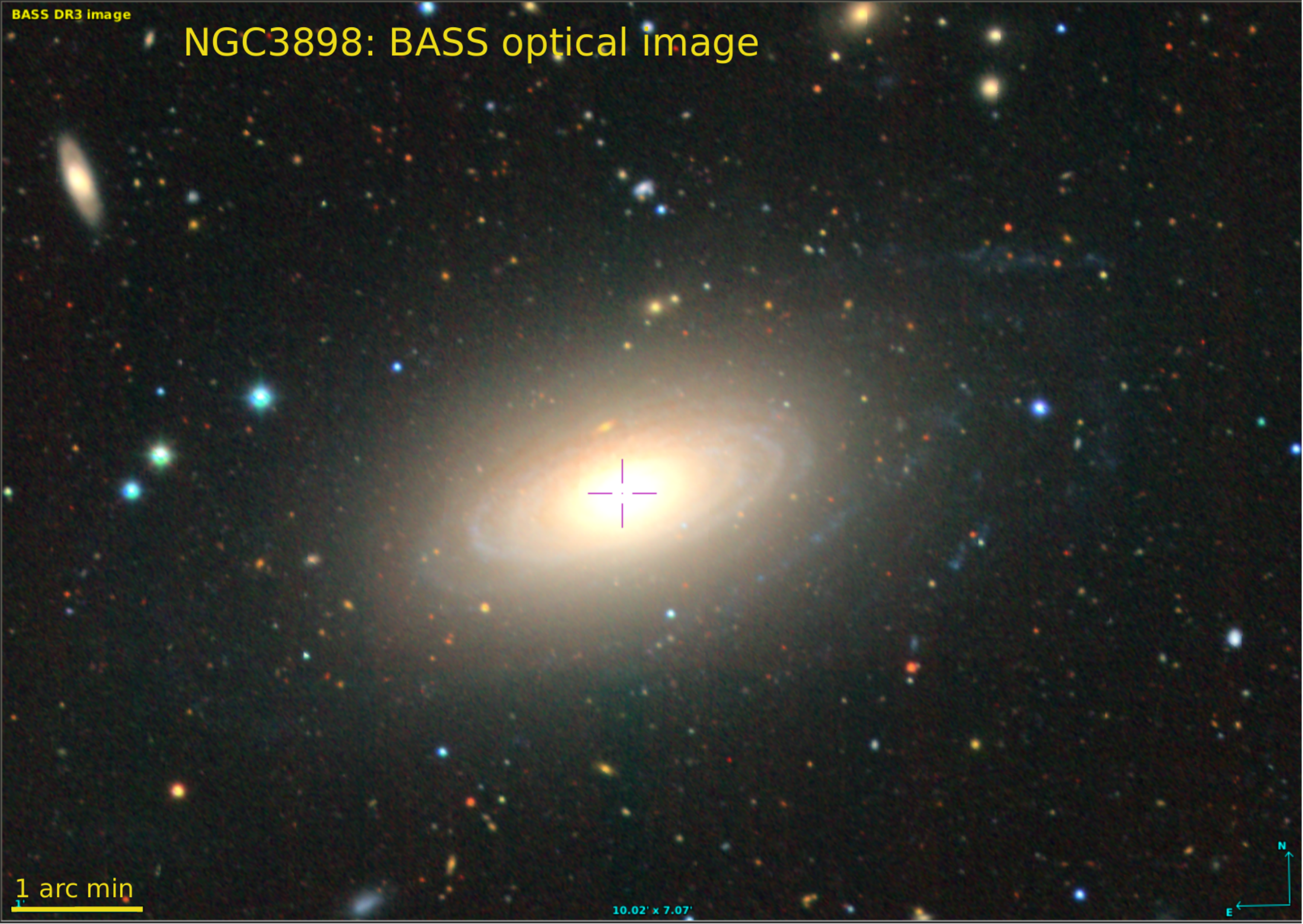}}
\subfloat[\centering]{\includegraphics[width=6.0cm]{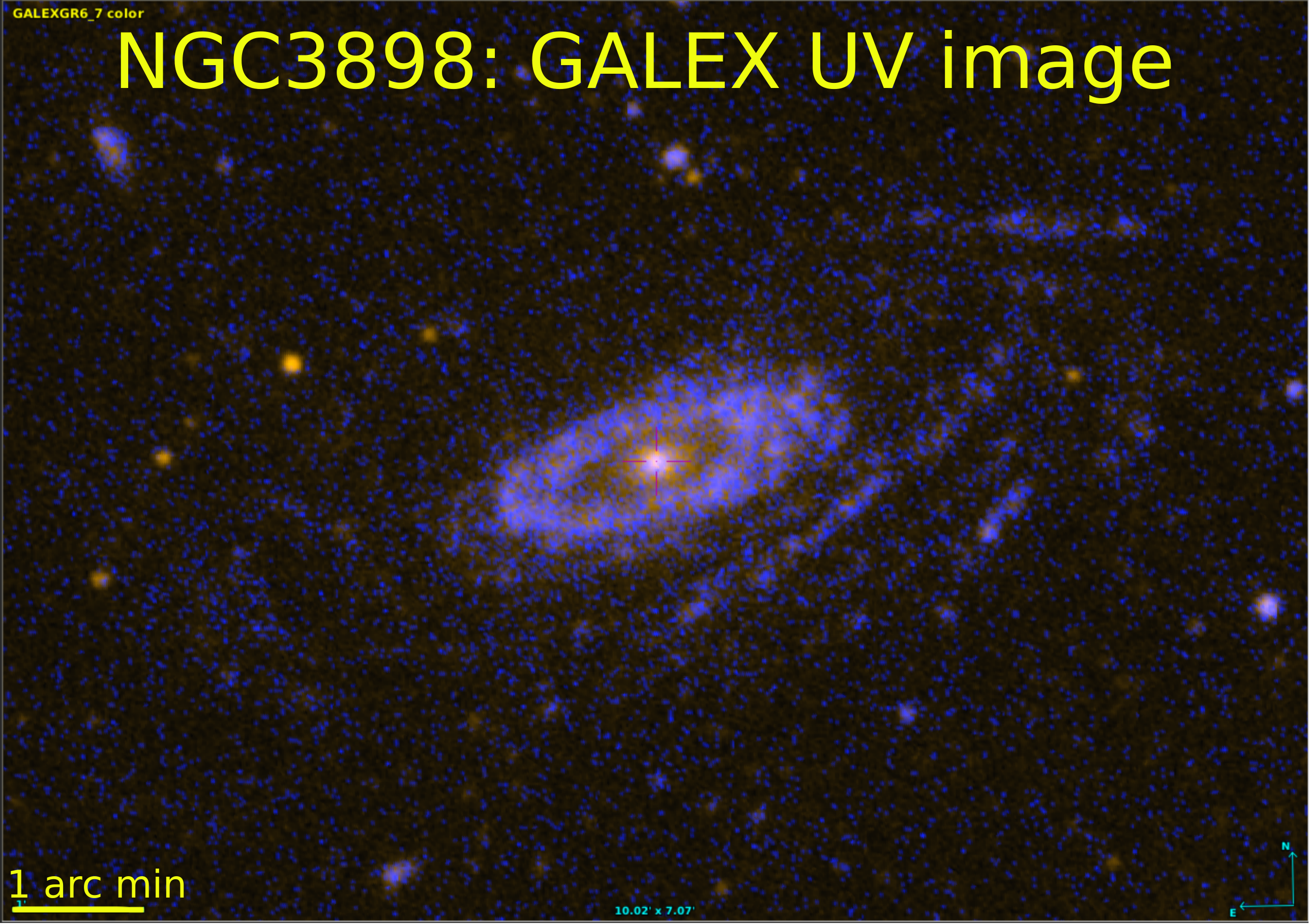}}\\
\end{adjustwidth}
\caption{ NGC3898: A deep multi-band optical colour image from BASS has been presented in the left panel. On the right panel, combined near and far UV images from GALEX have been presented. The scale bar in both images represents 1$^\prime$=4.86 kpc. The images are taken from Aladin Sky Atlas \citep{Aladin}.   
\label{fig:ngc3898_4}}
\end{figure} 
To investigate the regions where the radio lobe seems to touch the spiral-host galaxy for possible signs of jet-ISM interaction, we checked available images in GALEX (UV band) and deep optical images in BASS \citep{BASS}. Although we did not see any sign of jet-ISM interaction or signs of jet-triggered young star formation, we noticed that the western side of the galaxy shows extended young star-forming regions (Fig.~\ref{fig:ngc3898_4}). This is similar to the extended UV disks found in some galaxies earlier from GALEX UV studies \citep{Thilker2007}. These regions are also clearly seen, in BASS images, as arms/arcs in blue optical light. Note that they are still far from the western lobe to interact. 
\begin{figure}[H]
\begin{adjustwidth}{-\extralength}{0cm}
\centering
\subfloat[\centering]{\includegraphics[width=9.0cm]{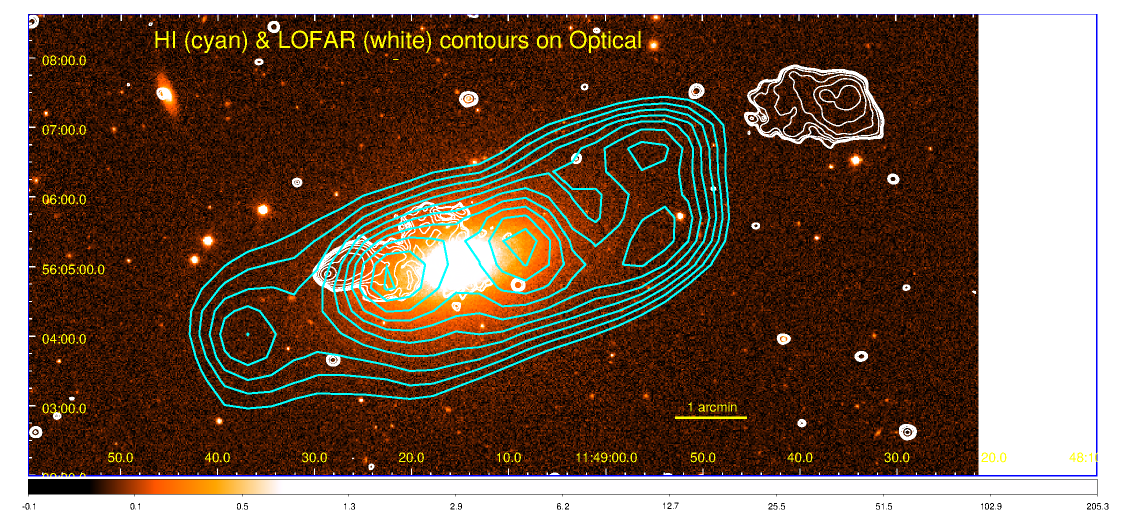}}
\subfloat[\centering]{\includegraphics[width=6.0cm]{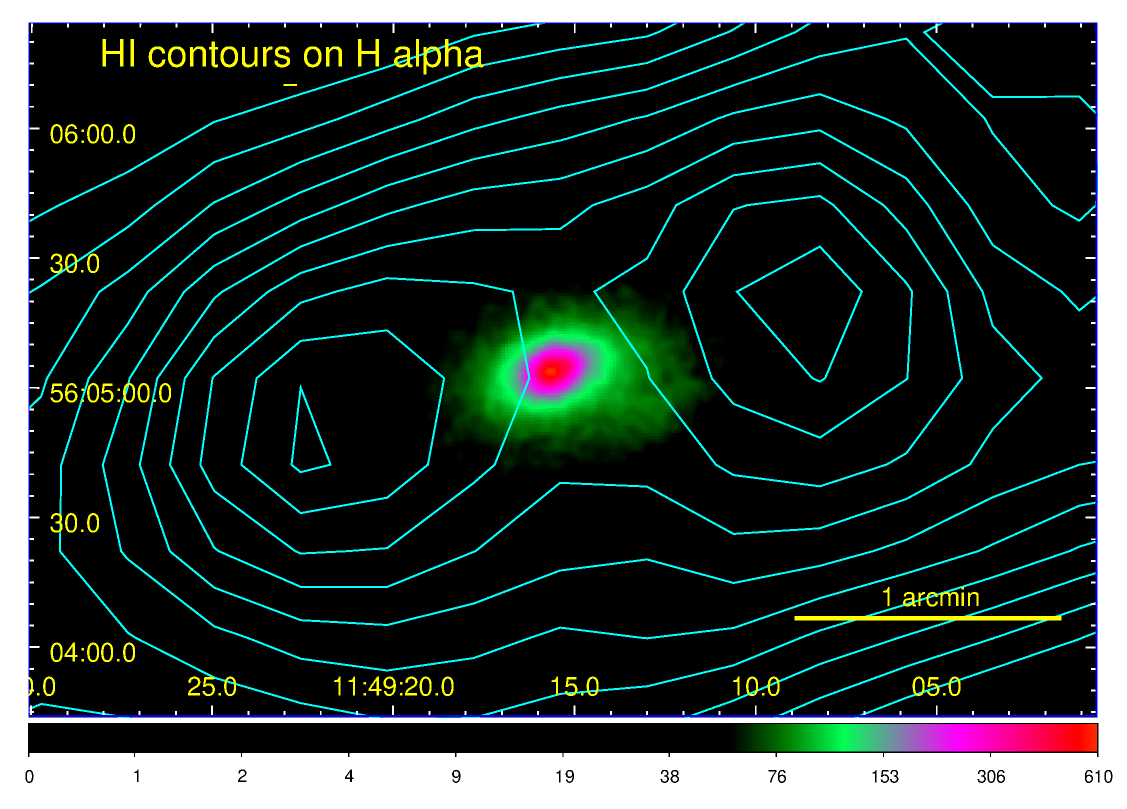}}\\
\end{adjustwidth}
\caption{ NGC3898: Total intensity  H\,{\sc i} 21cm line emission contours (cyan) have been overlaid on a r-band optical image of the galaxy from SDSS (left panel). LoTSS high-resolution radio contours (white) are also superposed in the same image. The same H\,{\sc i} contours but from the central region only are superposed on a false colour H$\alpha$ image (right panel). The scale bar for 1$^{\prime}$=4.86 kpc has been put at the right-hand bottom of both images. Both these  H\,{\sc i} and H$\alpha$ are available via the NASA Extragalactic Database (NED). \label{fig:halpahHI}}
\end{figure} 


We also looked for signatures of asymmetry in the gaseous medium.  Previous H\,{\sc i} and H$\alpha$ studies have found the galaxy to have an inclination angle of $\sim$54$^{\circ}$ with a smooth gaseous velocity field \citep{NGC3898HIHalpha, Pignatelli}. The  H\,{\sc i}  21cm line observation data were available from the Westerbork Synthesis Radio Telescope (WSRT). The total intensity  H\,{\sc i}-emission contours (cyan) are overlaid on the optical image from the Sloan Digital Sky Survey (SDSS) (Fig.~\ref{fig:halpahHI} left panel). The nuclear region shows a dip, the surrounding region shows two peaks, and the regions surrounding them show clear east-west asymmetry. While the eastern region shows a truncated/shrunken  H\,{\sc i}, the western emission is wider. In the same image, we have overlaid LoTSS high resolution contours (thin white). While the short eastern lobe is seen, in projection, inside the  H\,{\sc i} disk on the truncated side, the longer western lobe is seen outside the  H\,{\sc i} disk on the wider disk side.  We also inspected the H$\alpha$ line emission images to check for similar asymmetry. A false colour H$\alpha$ image of the central region of the galaxy is presented in the right panel of Fig.~\ref{fig:halpahHI}. Ionised gas distribution is also seen as asymmetric, with an extended gas tail seen to the south-west of the nuclear region. 
In summary, the radio lobe asymmetry, asymmetry in the atomic and ionised gas distribution and asymmetry in the star-forming regions in the extended disk of the host galaxy may have some causal connections. We shall return to this discussion in later sections.

\subsection{RAD-Thumbs up Galaxy: Speca or radio Phoenix?}

\begin{figure}[H]
\includegraphics[width=12 cm]{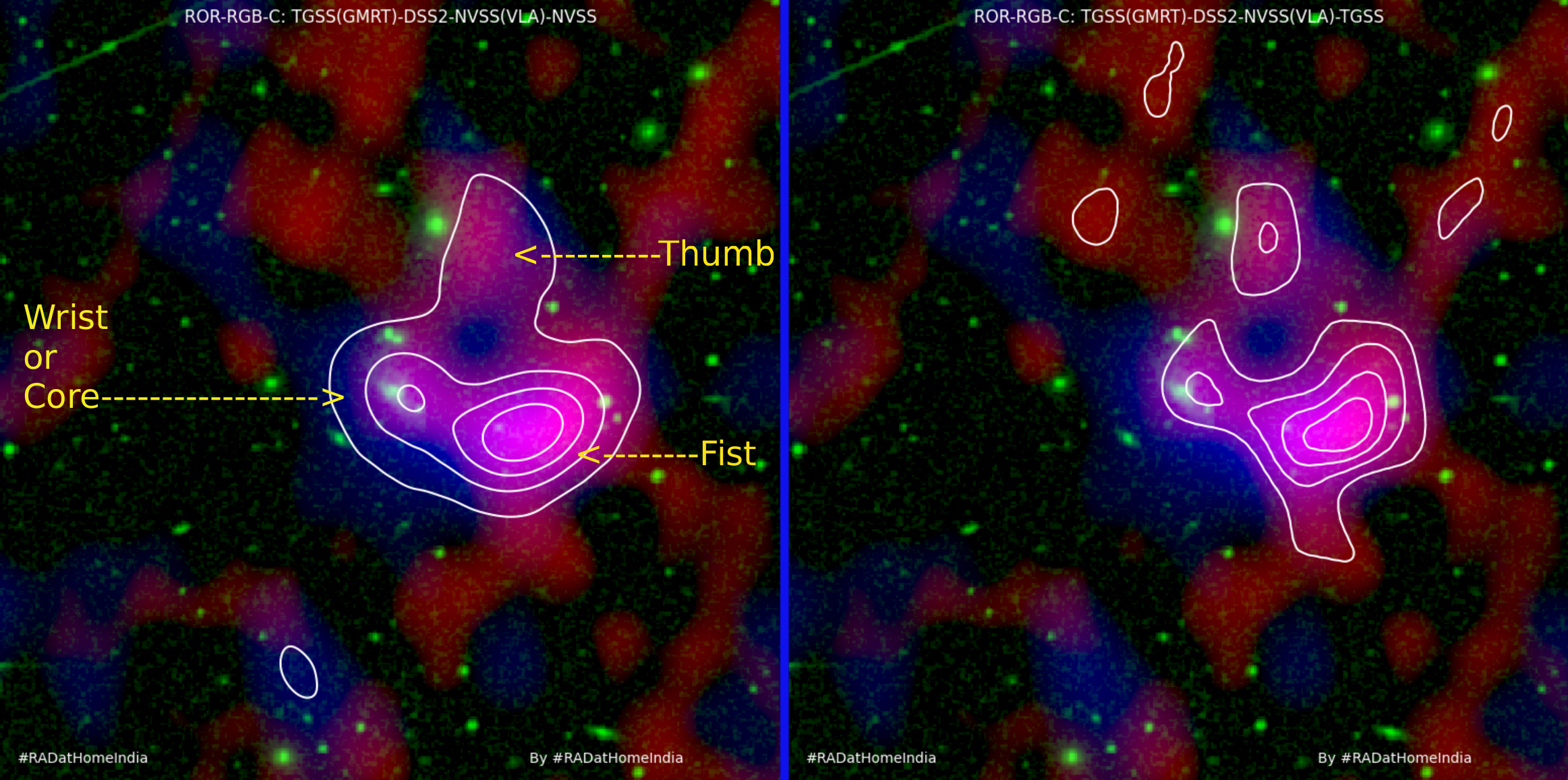}
\caption{RAD-``Thumbs up'' galaxy: As in Fig.~\ref{fig:ngc3898}, TGSS (red) DSS (green) and NVSS (blue) images of the target galaxy along with contours of NVSS on the left [0.0015, 0.004, 0.0065, 0.009 Jy beam$^{-1}$] and contours of TGSS on the right [0.015, 0.027, 0.038, 0.05 Jy beam$^{-1}$] are presented. The contours start with 3 times the r.m.s. noise in the respective radio images. NVSS has a resolution of b=45$^{\prime\prime}$ and TGSS has 25$^{\prime\prime}$ \label{fig:thumbsup1}.}
\end{figure}   

In December 2018, towards the end of a citizen science research training programme (a.k.a. RAD@home Discovery Camp), one participant spotted a radio source which seemed to be two radio lobes associated with a disturbed spiral galaxy but peculiarly showing emission on only one side of the host. As seen in Fig.~\ref{fig:thumbsup1}, there is no radio emission seen on the eastern side of the central optical galaxy. The radio features, seen in the NVSS image, looked like a ``Thumbs up'' sign and thus were discussed as the ``Thumbs up'' galaxy. The disturbed spiral galaxy, hereafter the host galaxy, can be identified with WISEA J221656.57-042434.1 ($z_{spec}=$0.095963 (scale =1.79 kpc/$^{\prime\prime}$ or 1$^{\prime}$=107 kpc).  There are three radio peaks which can be seen in both the NVSS and TGSS images. The brightest peak is away from the host galaxy on the west side, the secondary peak roughly coincides (shifted slightly to the west) with the host galaxy, and the faintest peak, the `thumb' feature, is seen on the northwest of the host galaxy.

\begin{figure}[H]
\begin{adjustwidth}{-\extralength}{0cm}
\centering
\subfloat[\centering]{\includegraphics[width=6.0cm]{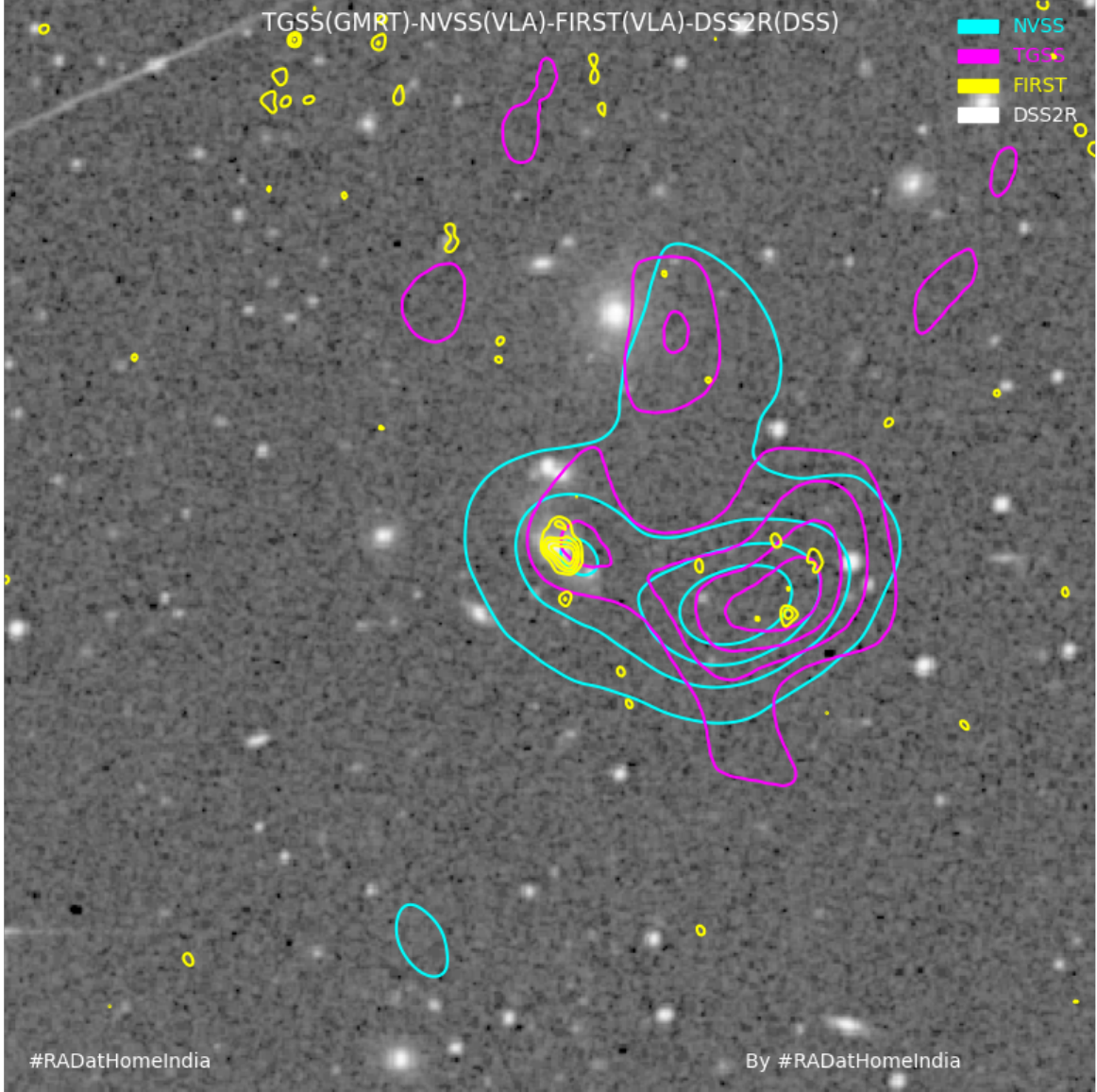}}
\subfloat[\centering]{\includegraphics[width=6.0cm]{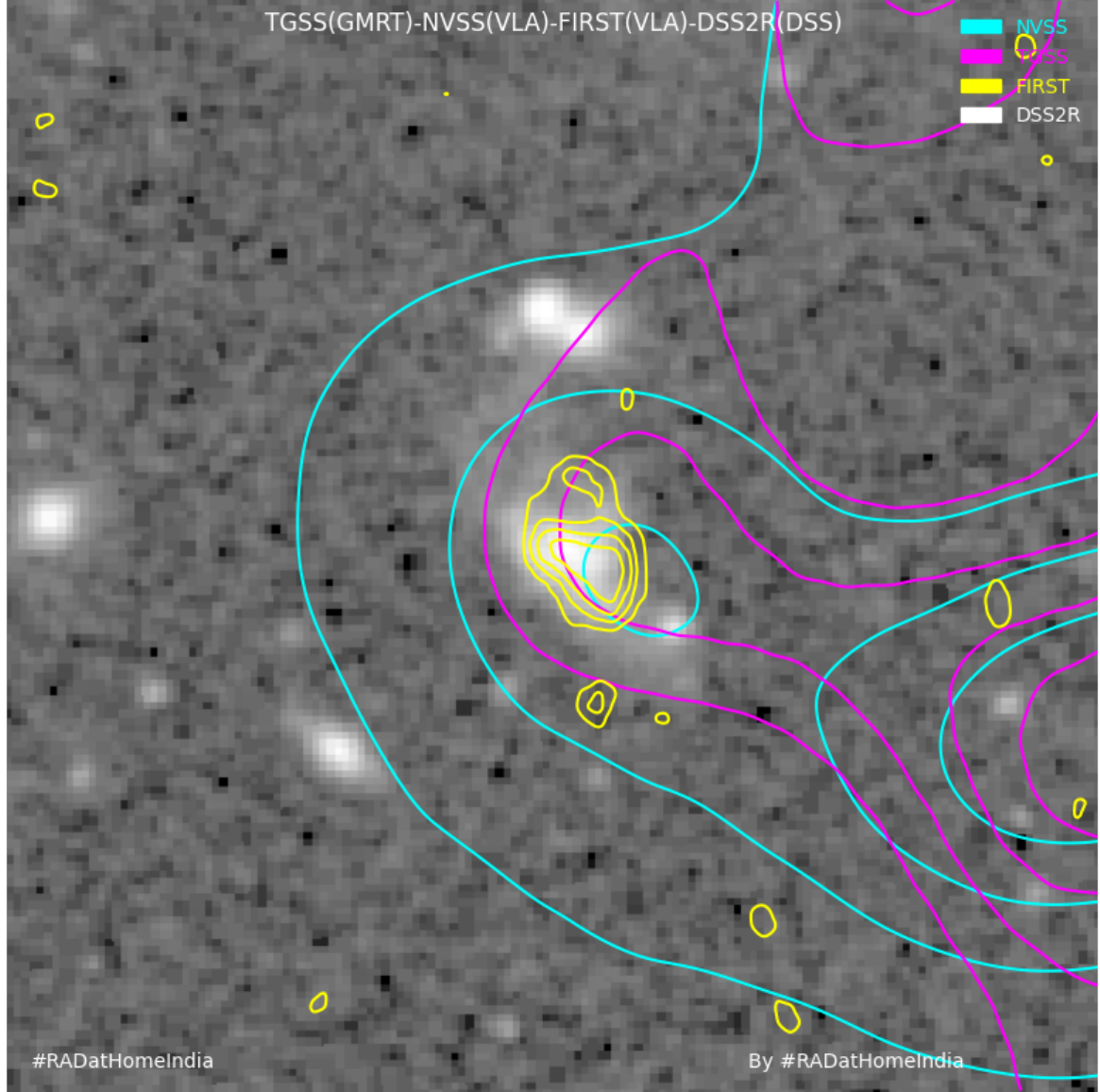}}\\
\end{adjustwidth}
\caption{ RAD-Thumbs up galaxy: [Left Panel] A typical RAD@home citizen science research web-tool output showing three radio data in contours on an optical (DSS-R) greyscale image. The contour levels for NVSS (cyan) are [0.0015 0.004 0.0065 0.009 Jy beam$^{-1}$],  for TGSS (magenta) are [0.015 0.027 0.038 0.05 Jy beam$^{-1}$], and for FIRST (yellow) are [0.0005 0.0007 0.0009 0.0011 Jy beam$^{-1}$]. The contours start with nearly 3 times the r.m.s. noise in the respective images. The resolution of the FIRST image is about 5$^{\prime\prime}$. Notice that despite being at the same 1400 MHz frequency and better rms noise sensitivity, FIRST is missing out on the diffuse emission (thumb and fist of the ``Thumbs up''). [Right panel] Contours and grey scale remain the same, but the image is a zoomed-in fraction of the left panel to show features around the host galaxy. \label{fig:thumbsup2}}
\end{figure} 

The FIRST survey data were available on the target. In Fig.~\ref{fig:thumbsup2} we present the RAD@home composite contour image of the target where the optical image is superposed with the same NVSS (cyan), TGSS (magenta) and FIRST (yellow) radio contours. FIRST, being a higher resolution (5$^{\prime\prime}$) image compared to NVSS (45$^{\prime\prime}$), it has resolved the compact peak at the host galaxy. However, the brightest NVSS/TGSS peak, the `fist' part, at the west of the host has been resolved out, which is evidence that it is unlikely to be a background radio source but a diffuse/remnant lobe possibly from the host galaxy. Similarly, the north-western peak, the `Thumb' part, has also been resolved out in the FIRST image.    

The right panel of the same figure shows a zoomed-in view of the same multi-contour image. The resolved nuclear peak shows a ``$<$" like structure. Its orientation and similarity with the larger-scale Thumbs-up structure are quite intriguing. The radio emission probably has a smaller, bent double-lobed structure. The western peak is comparatively brighter than the northwestern peak. This radio structure is similar in size to that of the host galaxy. Such sub-galaxy-scale radio lobes/jets are typical of jetted Seyfert galaxies. From the centre to the western peak of the mini-``Thumbs up'', the projected distance is 12$^{\prime\prime}$ (21.5 kpc) while that to the north-western peak is 10$^{\prime\prime}$ (17.9 kpc). In comparison, the nucleus to western radio peak of the large-scale ``Thumbs up'', or the fist, is 123$^{\prime\prime}$ (220 kpc). Similarly, from the nucleus to the peak at the Thumb is 101$^{\prime\prime}$ (181 kpc). Hence, in projection, the large radio lobes are at least ten times larger than the small-scale radio lobes. Since the radio jet ejection has to happen in opposite directions from the nucleus and both the small and large radio lobes are on the western side of the nucleus, we are likely seeing the source whose jet axis is either parallel to the line of sight or in the plane of the sky. Thus, assuming an intermediate inclination, the real linear size of the radio source can be bigger than 400 kpc (224$^{\prime\prime}$).      


\begin{figure}[H]
\includegraphics[width=9 cm]{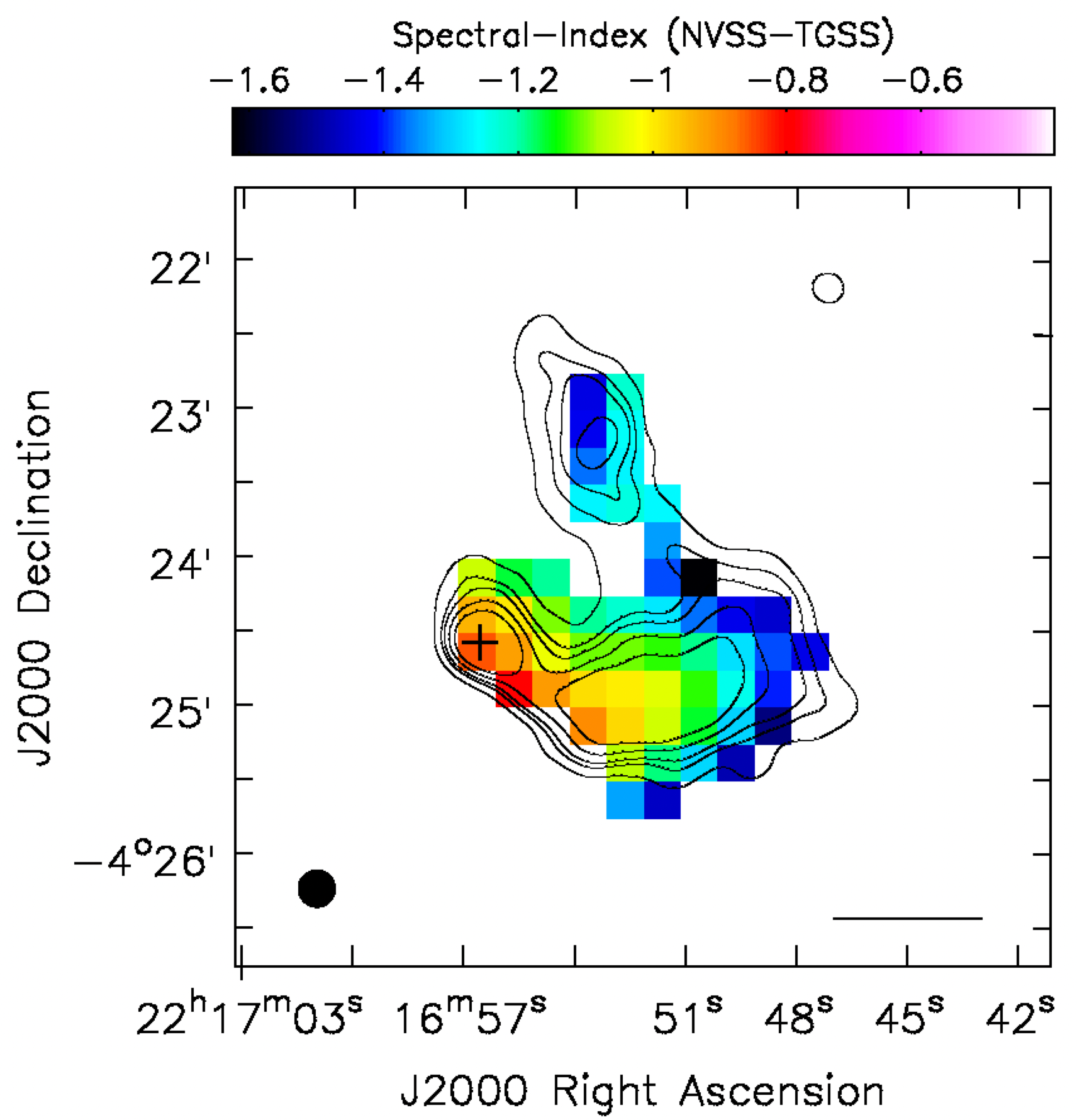}
\caption{ Spectral index ($\alpha_{150~\mathrm{MHz}}^{1400~\mathrm{MHz}}$) map of the RAD-``Thumbs up'' galaxy from the SPIDX database (NVSS+TGSS) overlaid with EMU 944 MHz radio contours (levels: rms$\times$[3, 6, 8, 12, 18], where rms $\sim$\,40 $\mu$Jy beam$^{-1}$) in black with its beam (15\arcsec) shown at the bottom left corner. The colour bar represents the respective spectral index values. The centre of the optical galaxy is shown with a "+" mark. Horizontal scale-bar at bottom right is of 1\arcmin=107 kpc length. \label{fig:thumbsup3}}
\end{figure}  


We obtained the spectral index map ($\alpha_{150~\mathrm{MHz}}^{1400~\mathrm{MHz}}$) of the  ``Thumbs Up'' radio source from the SPIDX database \citep{SPIDX}. This is presented in Fig.~\ref{fig:thumbsup3}, which clearly shows that the host galaxy region is typical of synchrotron radiation ($\alpha \sim$-0.75), and the diffuse region, farther away from the host, is $\alpha \sim$\, -1.5, significantly steeper. This suggests very little contribution of star formation for the mini-``Thumbs up'' and the presence of a small unresolved jetted AGN. This pattern is also consistent with expectations from the composite contour map, where NVSS peaks were resolved out in FIRST, suggesting the remnant or diffuse nature of the large-scale radio lobes.

\begin{figure}
\includegraphics[scale=0.23]{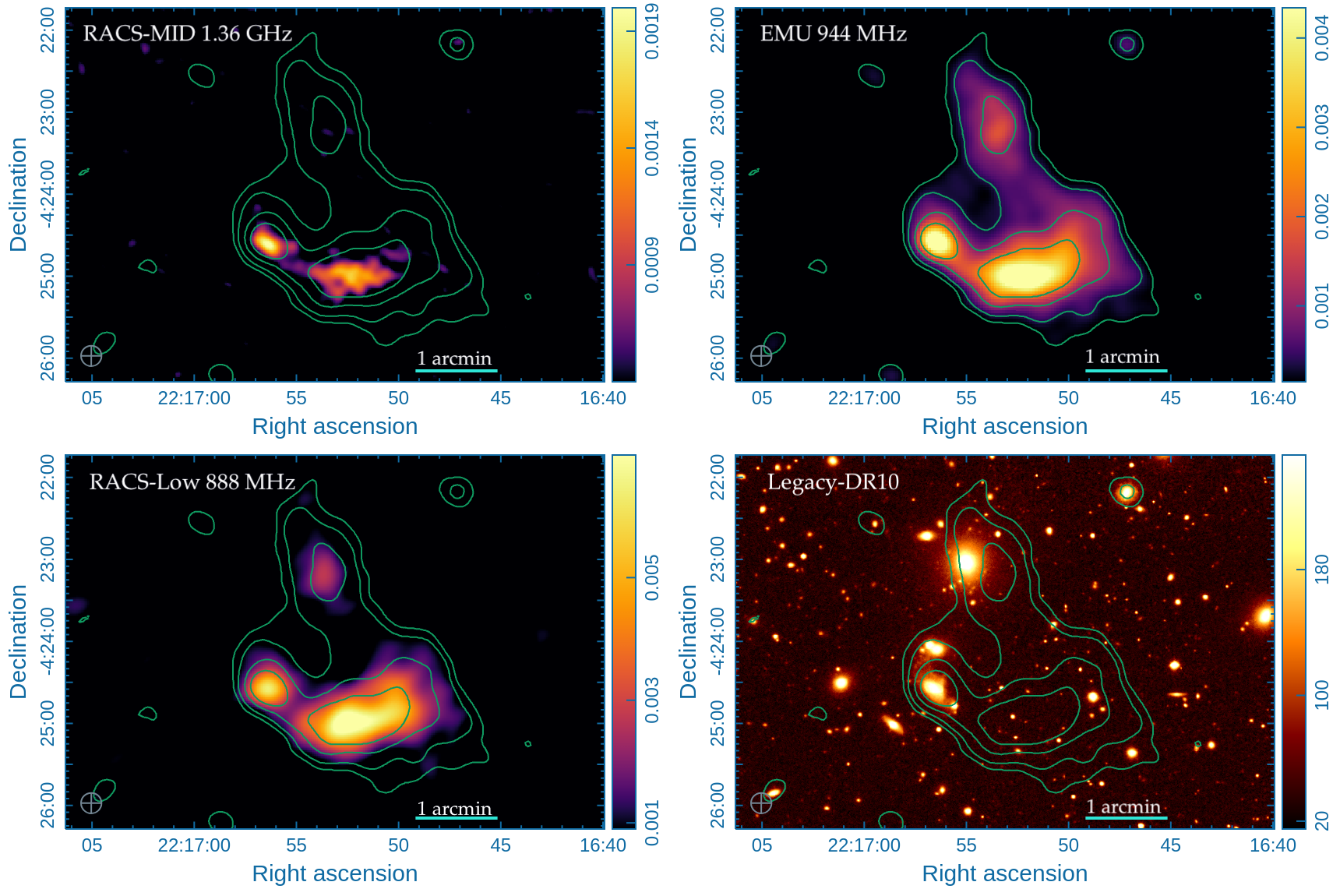}
\caption{Multi-wavelength view of the Thumbs Up galaxy. The top-left, top-right, and bottom-left panels show radio continuum images from RACS-MID (1.3 GHz), EMU (944 MHz), and RACS-LOW (888 MHz), respectively. The bottom-right panel presents the optical r-band image from the DESI Legacy Imaging Surveys DR10. All four panels are overlaid with green radio contours from the EMU 944 MHz image at levels of 0.14, 0.5, 1.5, 3.4, and 5.9 mJy beam$^{-1}$. Only emission above 3$\sigma$ is shown in the radio panels. The vertical colour bars indicate flux density in units of Jy beam$^{-1}$ for the radio images, and arbitrary units for the optical image. At the bottom-left corner, the EMU 944 MHz beam (15\arcsec) is shown as a grey circle with a central cross. Horizontal scale-bar at bottom right is of 1\arcmin=107 kpc length. \label{fig:thumbsup4}}
\end{figure}



Owing to the limited resolution and sensitivity of TGSS and NVSS, the thumb-like feature could not be reliably discerned or its morphology clearly characterised. To reassess the morphology with improved sensitivity and resolution, we examined recent data releases from the Rapid ASKAP Continuum Survey (RACS-low \& RACS-mid; \citep{RACS-low,RACS-mid}) and the Evolutionary Map of the Universe (EMU; \citep{EMU}). RACS-mid (Epoch~1, Stokes~I, 1367.5~MHz), RACS-low (Epoch~1, Stokes~I, 887.5~MHz) and EMU (943.491 MHz) images are shown in Fig.~\ref{fig:thumbsup4}.
The RACS-mid image reveals the radio core and a westward plume that extends toward the ``fist'' component of the ``Thumbs up''  structure. However, the ``thumb'' component is not detected, probably due to the reduced sensitivity to surface brightness at higher frequencies. In contrast, the RACS-low image captures the full extent of the ``Thumbs up''  morphology. While the bright ``fist'' remains connected to the radio core, the diffuse ``thumb'' appears detached, consistent with its low surface brightness and extended nature.

Fig.~\ref{fig:thumbsup4} also presents the EMU map of the ``Thumbs up''  galaxy with significantly higher sensitivity and clarity compared to other radio images. The EMU image has a rms noise level of 46\,$\mu$Jy\,beam$^{-1}$ and a spatial resolution of 15\arcsec. In addition to the previously identified core/wrist, fist, and thumb components, the improved sensitivity reveals filamentary connections between the thumb and the fist, as well as a faint one between the core and the thumb. The thumb component is now clearly resolved as an elongated structure oriented in the north-south direction. Its peak flux density is 1.8\,mJy\,beam$^{-1}$, corresponding to a signal-to-noise ratio of $\sim$\,39.
The detection of filamentary bridges and the extended plume from the core to the fist argues against a chance alignment of unrelated diffuse radio emission with the core component associated with the disturbed spiral host galaxy. 

The bottom-right panel of Fig.~\ref{fig:thumbsup4} shows the EMU radio contours overlaid on the $g$-band optical image from the Legacy Survey. As seen in earlier images, the radio core is co-spatial with the host galaxy. The fist has no radio peak corresponding to any optical galaxy, suggesting diffuse emission. The elongated thumb component exhibits a sharp eastern edge and does not show any associated radio emission from the nearby elliptical galaxy WISEA~J221654.97$-$042302.0 (also identified as WHL~J221654.9$-$042302 BCG, $z = 0.0869$), which is the brightest cluster galaxy (BCG) of the galaxy cluster WHL~J221654.9$-$042302 ($z_{sepc} = 0.098102$). This elliptical galaxy was also undetected in the FIRST radio survey image (Fig.~\ref{fig:thumbsup2}). 
In addition to this cluster at $z = 0.0869$ \citep{WenHan}, a supercluster has also been reported in the background at $z \sim 0.38$ \citep{Sankhyayan}. However, we detect no diffuse radio emission in the field other than that associated with the ``Thumbs up''  galaxy. The close redshift match suggests that the host galaxy (WISEA~J221656.57$-$042434.1 ($z = 0.095963$)) may be a member of the galaxy cluster WHL~J221654.9$-$042302, and as suggested by the asymmetric WAT-like radio lobes, it is likely circling the cluster BCG located to the east of the thumb. Hence, the large-scale diffuse thumb and fist components can possibly be part of the cluster diffuse emission \citep{vanWeeren2019}. The fist-thumb diffuse radio emissions are unlike cluster radio relics, which are typically seen in pairs, at the periphery of clusters, steep in spectral index and are associated with cluster-mergers. This is also unlike cluster radio halos, which are steep spectral diffuse sources and are seen around cluster BCGs without links to any individual galaxy. Hence, the thumb and fist seem similar to cluster radio phoenixes. A radio phoenix is a remnant radio lobe from past AGN-jet activity, but has recently been revived due to cluster-merger or similar such cluster dynamical processes \citep{EnsslinGopalKrishna, Mandal}. Since the disturbed spiral-host is the only possible host galaxy, seen in the middle, to have created these double remnant radio lobes, it is simply the remnant or older pair of lobes of the same host galaxy WISEA~J221656.57$-$042434.1. Hence, it is likely a case of a spiral-host episodic radio galaxy or Speca \citep{Hota2011} where the remnant lobes are misaligned and the host is a  WAT. Recent observations suggest that radio phoenixes are even possible in densities lower than clusters, that is, in group environments \citep{Riseley2025}. 


\begin{figure}[H]
\begin{adjustwidth}{-\extralength}{0cm}
\centering
\subfloat[\centering]{\includegraphics[width=6.0cm]{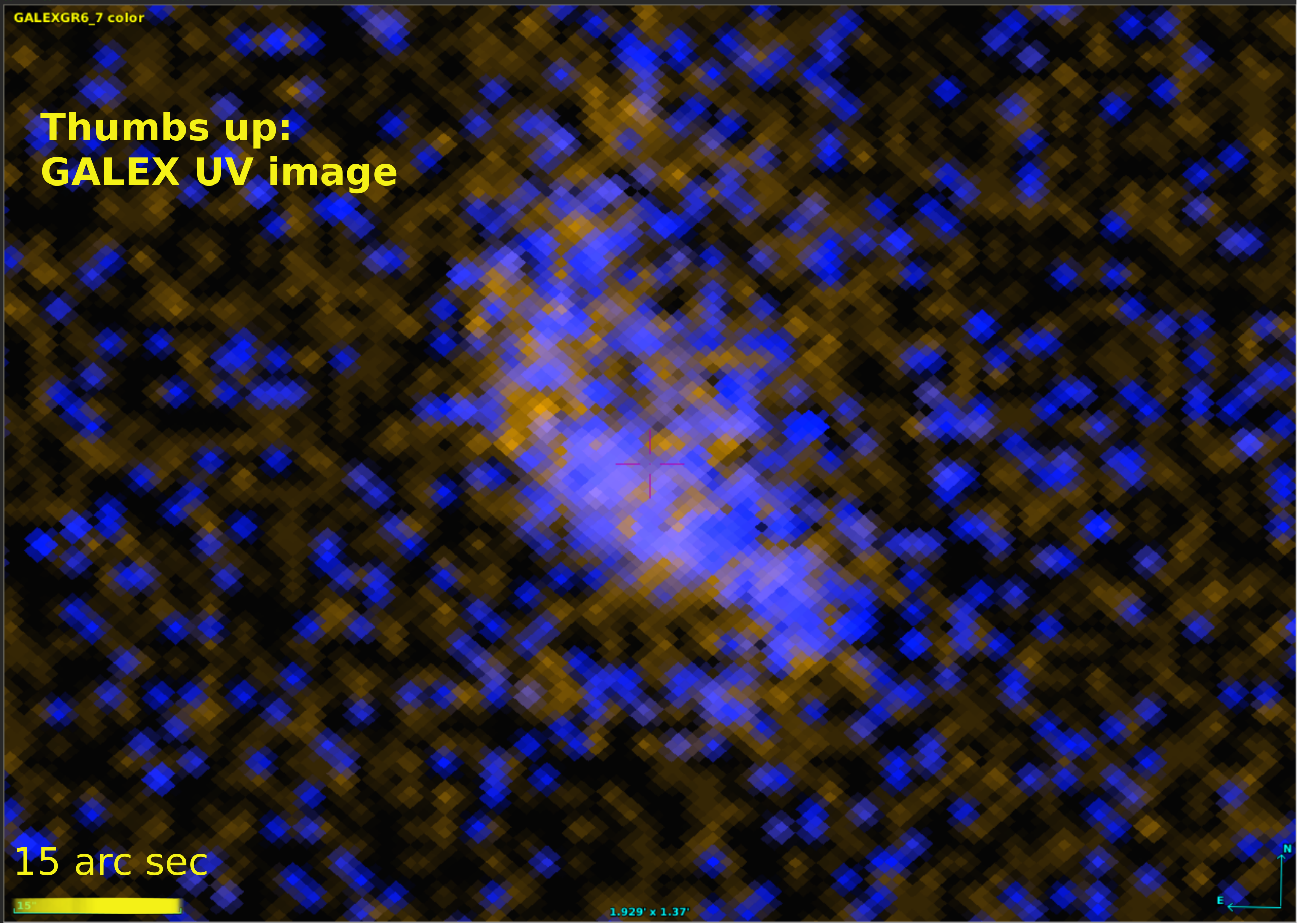}}
\subfloat[\centering]{\includegraphics[width=6.0cm]{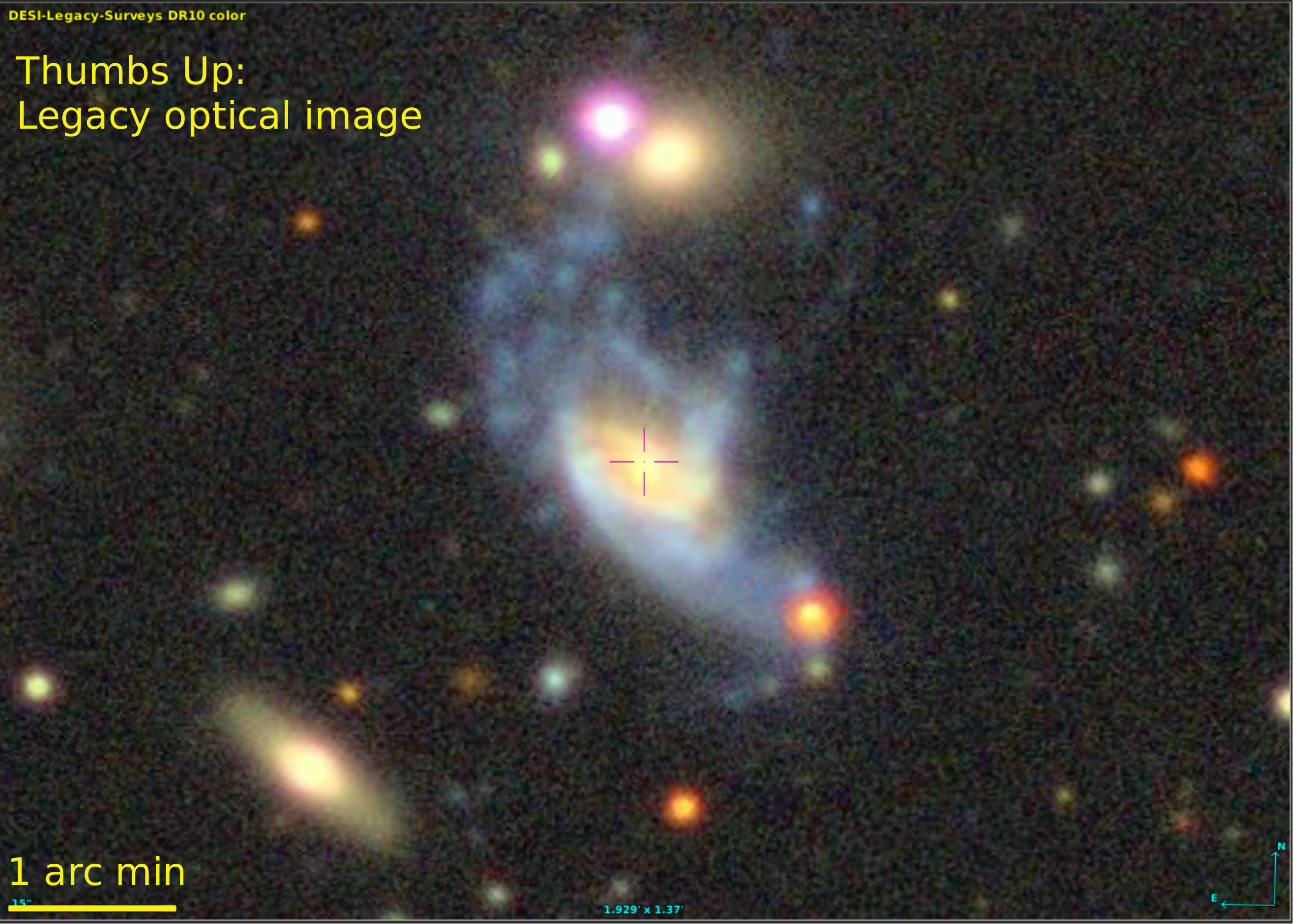}}\\
\end{adjustwidth}
\caption{ A combined near and far UV image of the RAD-``Thumbs up'' galaxy from GALEX (left panel). A deep optical colour image of the same galaxy from the Legacy survey. While the scale bar in the UV image (left) is 15$^{\prime\prime}$=26.85kpc that in the optical (right panel) image is 1$^\prime$=107 kpc. The images are taken from Aladin Sky Atlas \citep{Aladin}.
\label{fig:fig8}}
\end{figure} 

Ultraviolet images from GALEX and deeper optical images from the Legacy survey were available. Fig.~\ref{fig:fig8} (left panel) shows a combined near and far UV image of the host galaxy. The arc-shaped or bowl-shaped young star-forming region of the host galaxy naturally draws similarity with the ram-pressure-stripped galaxies in nearby clusters like Virgo and Coma. This UV image can be compared with GALEX images of Virgo cluster galaxies showing H\,{\sc i} evidence of ram pressure stripping \citep[e.g., NGC~4522 \& NGC~4330][]{Boissier}.  

Similarly, the Legacy optical image shows a disturbed spiral galaxy with bright blue star-forming clumps on its southern arm as well as disturbed regions on the north. The distribution of the star-forming regions of the galaxy has no correspondence with the sub-galactic mini-``Thumbs up'' radio structure seen in the FIRST image. This, along with the spectral index value, is a clear sign that the mini-``Thumbs up'' radio emission is due to AGN activity and not from star formation. The overall optical structure is also similar to that of the Virgo cluster galaxy NGC~4438. Interestingly, NGC~4438 also has a mini-radio double along with diffuse radio emission to the west of the galaxy, on a large scale, which is attributed to ram pressure stripping. A multi-wavelength study of NGC~4438 can be referred to for further comparisons \cite {Hota2007}. Although the arms are, the central region of the galaxy has not been disturbed. The host galaxy has a spectroscopic redshift $z=0.095963 \pm 0.000150$ or (v (heliocentric)=28769 $\pm$ 45 km s$^{-1}$). The early-type galaxy seen to the north is WISEA J221656.42-042405.6 which has a heliocentric recession velocity of 20870 km s$^{-1}$ (measurement uncertainty not available in NASA Extragalactic Database). Hence, these two galaxies, though they seem to be connected in projection, are widely separated in velocity space and are clearly not interacting. The late-type edge-on galaxy seen to the south-east of the host galaxy is undisturbed, and hence not part of any possible interaction. 

In summary, 
plume and filamentary links indicate that the compact ``mini‑thumbs up'' component is likely feeding the more diffuse fist and thumb features. Although the diffuse emission could, in principle, be a radio‑phoenix associated with the surrounding cluster, the simpler explanation is that both episodes of the radio lobes originate from the same spiral host and have possibly been pushed westward by ram pressure of the intra-cluster medium acting east to west. Similar environments have been reported for other spiral‑host radio galaxies \citep{Gao, Yuan}, and our results align with those studies.



\unskip


\section{Discussion}
Identifying the host galaxy of an extragalactic radio source can sometimes become challenging, particularly when the jet linking the radio core to the hotspots is absent. It is also complicated when the radio lobes are of non-standard shape or remnant radio lobes with diffuse emission blobs. In such cases, multiple potential host galaxies may lie near the radio structure, making the association ambiguous. In both systems presented in this study, the presence of a surrounding galaxy cluster/group further complicates host identification. In the case of NGC3898 the radio lobes, asymmetric to NGC3898, can be a symmetric background radio galaxy at a redshift near 0.469 or even at a higher redshift. 
Similarly, in the case of the ``Thumbs up''  galaxy, the core radio component is clearly associated with the disturbed spiral (WISEA J221656.57-197042434.1), but the 'fist' and 'thumb' components can be diffuse radio phoenix emission from some other host galaxy, but associated with the same cluster (WHL J221654.9-042302). 

Several mechanisms can contribute to radio lobe asymmetries, including: (1) light travel time effects, where the nearer lobe is observed at a later evolutionary stage compared to the lobe farther away; (2) Doppler boosting, wherein the approaching lobe appears brighter than the receding one; and (3) intrinsic asymmetries in the ambient medium on scales of several hundred kiloparsecs, leading to differential resistance to lobe expansion \citep{ODea2009, Gopal-Krishna}. Previous studies have tried to explain the correlated radio lobe asymmetries and optical asymmetries with motion of the host galaxy through the cluster medium, where the ram pressure stripped gas and dust causes the radio optical correlations \citep{GopalKrishnaWiita}. Multi-wavelength images of the host galaxy and the asymmetric radio lobes present observational evidence in support of such models involving ram pressure stripping.     
In the absence of X-ray data, we do not attempt to investigate the role of hot intracluster gas in detail. However, the UV and optical signatures in the host galaxies suggest the possible influence of ram pressure stripping in shaping the radio morphology. There may not be strong signs of ram pressure stripping like H\,{\sc i}/H$\alpha$ tails, but far from the stellar disk, where gravity is weak compared to ram pressure, the lobes may be bent/distorted relatively easily. Such effects are commonly observed in bent-double radio sources within group environments \citep{Freeland}.

\begin{figure}
    \centering
    \includegraphics[width=1\linewidth]{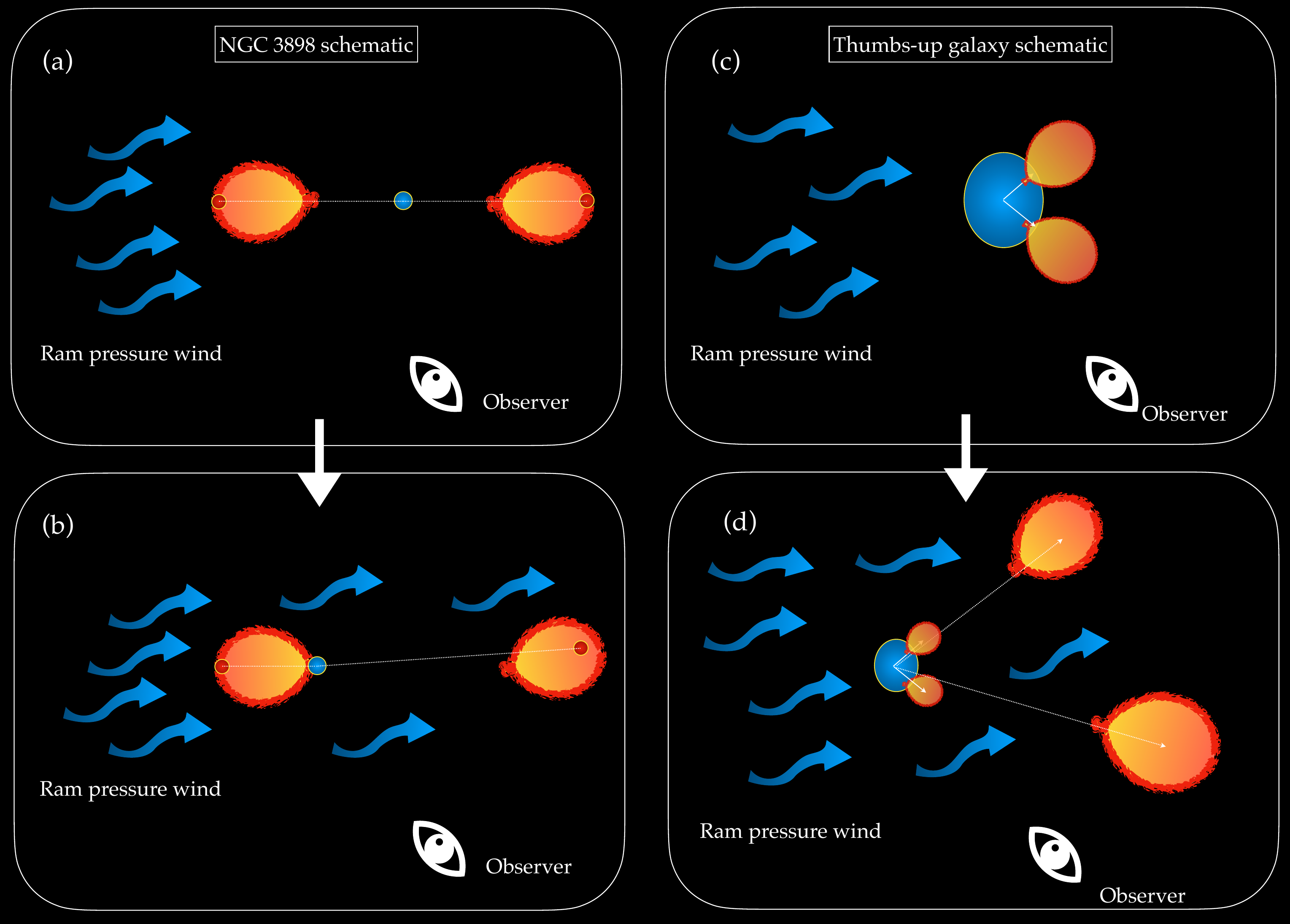}
    \caption{\textbf{Schematic illustration of the proposed ram-pressure evolution.}  
\emph{Left panels (a $\rightarrow$ b): NGC\,3898.}  
Ram pressure is assumed to act from east (left) to west (right).  In the initial state (a), the radio jets emerge symmetrically from the host (blue circle; stellar disc not to scale).  As the galaxy moves eastward through the intra-group medium, the eastern lobe encounters the oncoming wind, so its hotspot (red circle) remains close to the galaxy, while the back-flow of the western lobe is displaced beyond its hotspot (b).  The true three-dimensional jet orientation is uncertain, and intrinsic bending is omitted for clarity.
\emph{Right panels (c $\rightarrow$ d): RAD–``Thumbs Up'' galaxy.}  
The host (blue ellipse) lies on one side of two pairs of radio lobes, consistent with ram pressure acting from east to west.  In the earlier episode (c), the inner lobes (small orange ellipses) are close to the nucleus.  With time (d), the host drifts eastward and the older lobes expand (large orange ellipses), while a newer, more compact pair is launched.  The progressive westward offset of successive lobe generations reflects continued ram-pressure influence. }
    \label{fig:schematic}
\end{figure}


\subsection{NGC3898: not bending but stripping:} 
 Based on the available evidence (based on current data), we consider the asymmetric radio lobes to be most plausibly associated with NGC3898, which is a member of a group of galaxies. If the host galaxy moves roughly in the direction of the eastern radio jets, the hotspot will naturally be ahead of the bridge/back flow, and the core-to-hotspot distance will be small compared to that on the other side. On the other hand, the western side hotspot may be seen closer to the host than some of the diffuse emission in the lobe. While the jets/hotspots may be moving at a few percent of the speed of light, the host galaxies may be moving only at a few hundred km s$^{-1}$ in groups and to a couple of thousands km s$^{-1}$ in clusters. Such a group/cluster environment also creates head-tail and wide-angle-tailed (WAT) radio galaxies. By analogy with WAT radio galaxies, if the observed arm-length asymmetry in NGC~3898 were attributed to bending, one would expect the shorter (9~kpc) eastern lobe to be aligned closer to the line of sight, and the longer (28~kpc) western lobe to lie closer to the plane of the sky. However, the observed morphology contradicts this interpretation: the hotspot of the eastern (shorter) lobe lies at the farthest extent from the host galaxy, while in the western (longer) lobe, the backflow or diffuse emission is located beyond the hotspot. These structural details indicate that bending alone cannot account for the asymmetry, suggesting the influence of another force, such as ram pressure stripping. In optical and UV images, the eastern half of the host galaxy appears relatively clear, while the western half exhibits multiple star-forming clumps and extends further from the galactic centre. This asymmetry in the stellar disc supports an east-to-west ram pressure stripping scenario. Additional supporting evidence includes the observed depletion of H\,{\sc i} gas on the eastern side and the presence of an H$\alpha$ tail extending westward, both consistent with ram pressure acting from east to west. The stripping of the eastern radio lobe may also enable the backflow to reach or overtake the host galaxy. Similarly, on the western lobe, the ram pressure stripping can push the diffuse emission beyond the hotspot.  Notably, evidence of ram pressure stripping affecting both H\,{\sc i} and radio continuum emission has been reported in multiple group environments \citep{Ho124Kantharia,Freeland}, reinforcing the plausibility of such a mechanism in NGC~3898. As illustrated in the schematic (left panel a and b of Fig.~\ref{fig:schematic}), ram pressure stripping acting approximately along the jet axis can lead to two distinct effects: the backflow from the eastern lobe may reach or overlap with the host galaxy, while the lobe itself remains short; conversely, the backflow from the western lobe may be displaced beyond its hotspot, resulting in an elongated morphology. Gas stripping becomes effective only where the gravitational potential of the stellar disc is sufficiently weak to be overcome by the external ram pressure\citep{GunnStripping}. Consequently, the host galaxy often appears morphologically undisturbed, with regular gas kinematics. Only in deep H\,{\sc i} or H$\alpha$ imaging are long ($\sim$\,100--200~kpc) ram pressure-stripped tails typically revealed. Future deep H$\alpha$ observations may help confirm or rule out the proposed mechanism. Additionally, a spatially resolved radio spectral index map could serve as an independent diagnostic of ram pressure effects on the radio lobes.

\subsection{RAD-Thumbs Up galaxy: Bent, episodic, and stripped:}
When a FR I or FR II deviates from the standard structure and has diffuse emission due to the remnant radio lobes, it becomes difficult to associate it with a particular host galaxy. This becomes even more complicated in group/clusters of galaxies. The diffuse thumb and fist component can be debatable if it is associated with the disturbed spiral (WISEA J221656.57-197042434.1) or unrelated radio phoenix. Due to the plume connecting the core with the fist, a faint filament connecting the core to the thumb and an east-to-west gradient of the spectral index, we consider it as evidence that the disturbed spiral is the supplier of the diffuse plasma that is seen in the fist and thumb, which has aged with time. In some groups/clusters where jetted AGN is seen near diffuse emission, it has been suggested that the remnant radio lobes have probably been re-ignited by shocks and turbulence \citep{EnsslinGopalKrishna, Mandal, Riseley2025}. ``Thumbs up''  may be representing such a case of revived remnant radio lobes, known as a radio phoenix. 
The spectral index map and structure of the emission, compact vs diffuse emission, clearly suggest that the ``Thumbs up'' galaxy has two episodes of AGN-driven radio emission in the form of bent double lobes.  The right panel (c and d) of Fig.~\ref{fig:schematic} presents a schematic of the ``Thumbs up'' galaxy. Due to the ram pressure, both young and old pairs of lobes are seen to the west of the host galaxy. Remnant lobes (large orange ellipses) are located on the west, nearly 10 times away from the inner/younger lobes (small orange ellipses), and extend up to nearly 200 kpc from the center of the galaxy. As both of these lobes are on the same side of the host, the jet axis is expected to be inclined, between parallel to the line of sight and in the plane of the sky. Hence, the real linear size of this spiral-host episodic radio galaxy (Speca) may be larger than 400 kpc or larger than typical 200-300 kpc radio galaxies hosted in ellipticals. These lobes are unlike typical episodic FR-II or double-double radio galaxies \citep{SaikiaDDRG,sagan5}. Spectral index map of typical FR-II radio galaxies shows the backflow as a steep spectral index and the end of the lobe or hotspot regions as flat. If the lobe plasma was stripped away from the host, the plasma seen at the farthest point would be the oldest or steepest. This is what is seen in the present case, suggesting that ram pressure stripping has shaped the large-scale radio emission.    

\subsection{Unique Context:} 
The effects of ram pressure on WAT radio galaxies are commonly observed, but these systems are almost always hosted by ellipticals, where direct evidence of ram pressure stripping is difficult to trace due to the lack of gas and dust. In contrast, spiral galaxies often show clear signs of ram pressure stripping, yet they rarely host double-lobed radio structures that extend beyond the optical size of the galaxy. Some known examples of ram pressure-stripped spiral galaxies with double radio lobes, though small in extent, include NGC~4438 and NGC~4388 \citep{HummelSaikia}, and NGC~4569 \citep{NGC4569}.

Unless the radio lobes extend significantly beyond the host, it is not possible to directly study the effect of ram pressure from the ambient thermal gas on the nonthermal plasma of AGN-driven lobes. The two sources reported in this study provide rare examples where such interactions may be examined, offering unique targets for future observations with upcoming mega-telescopes. The interaction between thermal gas and nonthermal outflows has direct implications for the modelling of AGN feedback \citep{MukherjeeFeedback}, especially in understanding how the outflow energy couples with the surrounding gas in the interstellar medium (ISM) and circum-galactic medium (CGM), a process that remains poorly constrained.
Moreover, galaxies undergoing ram pressure stripping by intrafilament gas flows around clusters can also serve as useful tracers to constrain the properties of diffuse gas in the cosmic web.

\vspace{6pt} 





\authorcontributions{
PA, AH and PD have contributed significantly to the manuscript writing. AH conceptualised the project. PA, DN, and PKN are trained citizen scientists involved in object discovery. AP has contributed significantly towards the overall success of the project from inception to completion.  
}

\funding{Results reported in this paper have been obtained from RAD@home citizen science research. RAD@home has not received any funding. However, the names of two institutions that have contributed by hosting RAD@home Discovery Camps, which have helped this particular paper, are the International Center for Theoretical Sciences of the Tata Institute of Fundamental Research (Code: ICTS/RADatICTS2018/05) and Nehru Planetarium, Prime Ministers Museum and Library, Ministry of Culture, Govt of India. 
}

\institutionalreview{
“Not applicable” 
}

\informedconsent{
``Not applicable'' 
}

\dataavailability{All the data used in this research are publicly available and have been cited as per the common practice. 
}

\acknowledgments{
We thank the referees for their critical comments, which have significantly improved the paper. PA is grateful to Prof. K. S. Kiran for his guidance during this project. AH acknowledges the University Grants Commission (UGC, Ministry of Education, Government of India) for his monthly salary since June 2014. We express our profound gratitude to the Late Dr. Nandivada Rathnasree, who was instrumental in organising multiple RAD@home Discovery Camps/workshops at Nehru Planetarium (Delhi). The long list of national and international organizations which have helped the establishment and growth of the first Indian astronomical citizen science research platform, RAD@home, are acknowledged in detail at \url{https://radathomeindia.org/brochure}.
We thank the staff of the GMRT who made these observations possible. GMRT is run by the National Centre for Radio Astrophysics of the Tata Institute of Fundamental Research (India). This scientific work uses data obtained from Inyarrimanha Ilgari Bundara / the Murchison Radio-astronomy Observatory. We acknowledge the Wajarri Yamaji People as the Traditional Owners and native title holders of the Observatory site. CSIRO’s ASKAP radio telescope is part of the Australia Telescope National Facility (\url{https://ror.org/05qajvd42}). Operation of ASKAP is funded by the Australian Government with support from the National Collaborative Research Infrastructure Strategy. ASKAP uses the resources of the Pawsey Supercomputing Research Centre. Establishment of ASKAP, Inyarrimanha Ilgari Bundara, the CSIRO Murchison Radio-astronomy Observatory and the Pawsey Supercomputing Research Centre are initiatives of the Australian Government, with support from the Government of Western Australia and the Science and Industry Endowment Fund. This paper includes archived data obtained through the CSIRO ASKAP Science Data Archive, CASDA (\url{https://data.csiro.au}).
}

\conflictsofinterest{
The authors declare no conflicts of interest.
} 




\begin{adjustwidth}{-\extralength}{0cm}

\reftitle{References}



\begin{thebibliography}{999}
\bibitem[]{Hota2011}
Hota, A; Sirothia, S.K.;  Ohyama, Y. et al., Discovery of a spiral-host episodic radio galaxy, {\em MMRAS} {\bf 2011}, {\em 417},  L36--L40.
\bibitem[]{Bagchi2014}
Bagchi, J.; Vivek, M.; Vikram, V. et al. Megaparsec relativistic jets launched from an accreting supermassive black hole in an extreme spiral galaxy, {\em ApJ} {\bf 2014} {\em 788} 174-191 
\bibitem[]{Sethi2025}
Sethi, S.; Kuźmicz, A.; Hunik, D.; Jamrozy, M., Serendipitous discovery of a spiral host in a 2 Mpc double-double lobed radio galaxy, {\em A\&A}, {\bf 2025}, {\em (in press)}

\bibitem[]{Baum}
Baum, S. A.;  O'Dea, C. P.;  Dallacassa, D.; de Bruyn, A. G.; Pedlar, A., Kiloparsec-Scale Radio Emission in Seyfert Galaxies: Evidence for Starburst-driven Superwinds?
{\em ApJ} {\bf 1993} {\em 419}, 553-572

\bibitem[]{Colbert}
Colbert E. J. M.; Baum S. A.; Gallimore J. F.; O’Dea C. P.; Christensen J.
A.,  Large-Scale Outflows in Edge-on Seyfert Galaxies. II. Kiloparsec-Scale Radio Continuum Emission, {\em ApJ}, {\bf 1996}, {\em 467}, 551-578

\bibitem[]{Gallimore}
Gallimore, Jack F. ;  Axon, David J. ;  O'Dea, Christopher P. ;  Baum, Stefi A. ;  Pedlar, Alan, A Survey of Kiloparsec-Scale Radio Outflows in Radio-Quiet Active Galactic Nuclei, {\em AJ}, {\bf 2006}, {\em 132}, 546–569

\bibitem[]{Hota2006}
Hota A.  \& Saikia D.J., Radio bubbles in the composite AGN-starburst galaxy NGC 6764, 
{\em MNRAS}, {\bf 2006},  {\em 371}, 945–956

\bibitem[]{Webster2021}
Webster, B.; Croston, J.H.; Mingo, B.;  et al. , A population of galaxy-scale jets discovered using LOFAR, {\em MNRAS}, {\bf 2021}, {\em 500}, 4921–4936

\bibitem[]{Cecil}
Cecil, G.; Greenhill, L.J.; DePree, C.G. et al., The Active Jet in NGC 4258 and Its Associated Shocks, {\em ApJ}, {\bf 2000}, {\em 536},  675-696 

\bibitem[]{Capetti}
Capetti, A.; Axon, D. J.;  Macchetto, F. D.; Marconi, A.;  Winge, C., The Origin of the Narrow-Line Region of Markarian 3: An Overpressured Jet Cocoon, {\em ApJ}, {\bf 1999}, {\em 516}, 187194

\bibitem[]{VerdoesKleijn}
Verdoes Kleijn, G.A.  \& de Zeeuw, P.T., A dichotomy in the orientation of dust and radio jets in nearby low-power radio galaxies, {\em A\&A}, {\bf 2005}, {\em 435}, 43–64
\bibitem[]{Kinney}
Kinney, A. L. ;  Schmitt, H. R. ;  Clarke, C. J. ;  Pringle, J. E. ;  Ulvestad, J. S. ;  Antonucci, R. R. J., Jet Directions in Seyfert Galaxies, {\em ApJ}, {\bf 2000}, {\em 537}, 152-177

\bibitem[]{GunnStripping}
Gunn, James E. ;  Gott, J. Richard, III On the Infall of Matter Into Clusters of Galaxies and Some Effects on Their Evolution,  {\em APJ}, {\bf 1972}, {\em 176}, 1

\bibitem[]{AbadiStripping}
Abadi, Mario G.  ; Moore, Ben ; Bower, Richard G, Ram pressure stripping of spiral galaxies in clusters, {\em MNRAS}, {\bf 1999}, {\em 308}, 4,  947-954

\bibitem[]{Ignesti}
Ignesti, A. et al. Walk on the Low Side: LOFAR Explores the Low-frequency Radio Emission of GASP Jellyfish Galaxies, {\em ApJ}, {\bf 2022}, 937, 2, 58, 16

\bibitem[]{Croton}
Croton, D., et al.,  {\em MNRAS}, {\bf 2006}, {\em 365}, 111.
\bibitem[]{Hardcastle}
Hardcastle, M. J.;  Croston, J. H., Radio galaxies and feedback from AGN jets, {\em NewAR}, {\bf 2020}, {\em 88}, 101539

\bibitem[]{Hota2016}
Hota, A.; Konar, C.; Stalin, C.S. et al. Tracking Galaxy Evolution Through Low-Frequency Radio Continuum Observations using SKA and Citizen-Science Research using Multi-Wavelength Data, {\em J Astrophys Astron}, {\bf 2016},  {\em 37}, 41-

\bibitem[]{Hota2022}
Hota, A; Dabhade, P.; Vaddi, S.;  Konar, C.; Pal, S.;  Gulati, M.; Stalin, C.S.; Avinash, Ck.; Kumar, A.;  Rajoria, M.; Purohit, A. RAD@home citizen science discovery of an active galactic nucleus spewing a large unipolar radio bubble on to its merging companion galaxy, {\em MNRAS}, {\bf 2022}, {\em 517}, L86–L91


\bibitem[]{WISE}
Wright, E.L. et al., The Wide-field Infrared Survey Explorer (WISE): Mission Description and Initial On-orbit Performance, {\em AJ}, {\bf 2010}, {\em 140}, 6, 1868-1881

\bibitem[]{GALEX}
Morrissey, A. et al., The Calibration and Data Products of GALEX, {\em ApJS}, {\bf 2007}, {\em 173}, 682

\bibitem[]{Kumar2021}
Kumar, A; Avinash Ck.; Purohit A.; Hota A., RAD@home RGB-maker web-tool for citizen science research in multi-wavelength study of AGNs with radio jets. {\em Proceedings of the International Astronomical Union} {\bf 2021}, {\em 17(S375)}, 40-41

\bibitem[]{TGSS}
Intema, H. T. ; Jagannathan, P. ;  Mooley, K. P.  ;  Frail, D. A.,  The GMRT 150 MHz all-sky radio survey. First alternative data release TGSS ADR1, {\em A\&A}, {\bf 2017},  {\em 598}, A78

\bibitem[]{NVSS}
Condon, J. J. ; Cotton, W. D. ; Greisen, E. W. ; Yin, Q. F. ; Perley, R. A. ; Taylor, G. B.; Broderick, J. J. The NRAO VLA Sky Survey,  {\em AJ}, {\bf 1998}, {\em 115}, 1693-1716.


\bibitem[]{FIRST}
Becker, Robert H. ;  White, Richard L. ;  Helfand, David J., The FIRST Survey: Faint Images of the Radio Sky at Twenty Centimeters, {\em AJ}, {\bf 1995}, {\em 450}, 559-

\bibitem[]{SPIDX}
de Gasperin, F. ;  Intema, H. T. ;  Frail, D. A., A radio spectral index map and catalogue at 147-1400 MHz covering 80 per cent of the sky, {\em MNRAS}, {\bf 2018}, {\em 474}, 4, 5008-5022

\bibitem[]{NGC3898group}
Trentham, Neil ; Tully, R. Brent ; Verheijen, Marc A. W., The Ursa Major cluster of galaxies - III. Optical observations of dwarf galaxies and the luminosity function down to M$_R$=-11, {\em MNRAS},{\bf 2001} , {\em 325}, 1, 385-404.

\bibitem[]{Tempel2017}
Tempel, E ; Tuvikene, T ; Kipper, R ; \& Libeskind, N. I., Merging groups and clusters of galaxies from the SDSS data-The catalogue of groups and potentially merging systems, {\em A\&A}, {\bf 2017}, {\em 602}, A100.


\bibitem[]{LoTSS}
Shimwell, T. W. ;  Hardcastle, M. J. ; Tasse, C. ; Best, P. N. ; Röttgering, H. J. A.  et al. The LOFAR Two-metre Sky Survey. V. Second data release, {\em A\&A}, {\bf 2022}, {\em 659}, 27

\bibitem[]{Legacy}
Dey, Arjun et al. Overview of the DESI Legacy Imaging Surveys {\em The Astronomical Journal}, {\bf 2019}, 157, 5, 168, 29

\bibitem[]{BASS}
Zou, H.  et al. Project Overview of the Beijing-Arizona Sky Survey, {\em Publications of the Astronomical Society of the Pacific}, {\bf 2017}, {\em 129}, 976,  064101

\bibitem[]{LOFARaccuracyShimwell}
Shimwell, T. W. et al. The LOFAR Two-metre Sky Survey. II. First data release, {\em A\&A}, {\bf 2019}, {\em 622}, 1

\bibitem[]{GRSREV}
Dabhade, Pratik. ; D. J. Saikia. ;  and Mahato, Mousumi. Decoding the giant extragalactic radio sources, {\em Journal of Astrophysics and Astronomy}, {\bf 2023}, {\em 44.1}, 13.

\bibitem[]{Aladin}
Bonnarel, F.  et al. The ALADIN interactive sky atlas. A reference tool for identification of astronomical sources {\em Astronomy and Astrophysics Supplement}, {\bf 2000}, {\em 143}, 33-40


\bibitem[]{Thilker2007}
David A. Thilker et al., A Search for Extended Ultraviolet Disk (XUV-Disk) Galaxies in the Local Universe, {\ bf 2007}, {\em ApJS}, 173 538

\bibitem{NGC3898HIHalpha}
Vega Beltrán, J. C. ; Pizzella, A.; Corsini, E. M.; Funes, J. G. ; Zeilinger, W. W. ; Beckman, J. E. ; Bertola, F., Kinematic properties of gas and stars in 20 disc galaxies, {\bf 2001}, {\em A\&A}, 374, 394-411

\bibitem{Pignatelli}
E. Pignatelli, E. M. et al., Modelling gaseous and stellar kinematics in the disc galaxies NGC 772, 3898 and 7782, {\bf 2001}, {\em MNRAS}, {\em 323}, 1, 188–210

\bibitem[]{RACS-low}
McConnell, D. et al., The Rapid ASKAP Continuum Survey I: Design and first results, {\em Publications of the Astronomical Society of Australia}, {\bf 2020}, {\em 37}, 48

\bibitem[]{RACS-mid}
Duchesne, S. W., et al., The Rapid ASKAP Continuum Survey IV: continuum imaging at 1367.5 MHz and the first data release of RACS-mid, {\em Publications of the Astronomical Society of Australia}, {\bf 2023}, {\em 40}, e034.


\bibitem[]{EMU}
Hopkins, A. M., et al. The Evolutionary Map of the Universe: A new radio atlas for the southern hemisphere sky. {\em Publications of the Astronomical Society of Australia}, \textbf{2025}, {\em 1-32}.

\bibitem[]{WenHan}
Wen, Z. L. and Han, J. L., Calibration of the Optical Mass Proxy for Clusters of Galaxies and an Update of the WHL12 Cluster Catalog, {\bf 2015}, {\em ApJ}, 807, 178

\bibitem[]{Sankhyayan}
Sankhyayan, Shishir et al., Identification of Superclusters and Their Properties in the Sloan Digital Sky Survey Using the WHL Cluster Catalog, {\bf 2023}, {\em ApJ}, 958, 1, 62, 19

\bibitem[]{vanWeeren2019}
van Weeren, R. J.; de Gasperin, F.; Akamatsu, H.  et al.,  Diffuse Radio Emission from Galaxy Clusters, {\em Space Science Reviews}, {bf 2019}, 215, 1, 16, 75 

\bibitem[]{EnsslinGopalKrishna}
Enßlin, T. A.; Gopal-Krishna, Reviving fossil radio plasma in clusters of galaxies by adiabatic compression in environmental shock waves, {\em A\&A}, {\bf 2001}, 366, 26-34

\bibitem[]{Mandal}
Mandal, S. ; Intema, H. T. ; van Weeren, R. J.  et al. Revived fossil plasma sources in galaxy clusters, {\em A\&A}, {\bf 2020}, 634, A4, 11

\bibitem[]{Riseley2025}
Riseley, C. J. et al. Relighting the fire in Hickson Compact Group (HCG) 15: Magnetised fossil plasma revealed by the SKA Pathfinders and Precursors, {\em A\&}, {\bf 2025}, {\em 697}, 45, 32

\bibitem[]{Wen}
Wen, Z. L. ; Han, J. L.; Liu, F. S., A Catalog of 132,684 Clusters of Galaxies Identified from Sloan Digital Sky Survey III, {\em The Astrophysical Journal Supplement}, {\bf 2012}, {\em 199},  2, 34, 12

\bibitem[]{Boissier}
Boissier, S. ;  Boselli, A. ; Duc, P. -A.  et al. The GALEX Ultraviolet Virgo Cluster Survey (GUViCS). II. Constraints on star formation in ram-pressure stripped gas, {\em A\&A}, {\bf 2012}, {\em 545}, A142, 16   

\bibitem[]{Hota2007}
Hota, A; Saikia, D.J.; Irwin, J.A.,  NGC 4438 and its environment at radio wavelengths, {\em MNRAS}, {\bf 2007}, {\em 380}, 3, 1009–1022

\bibitem[]{Gao}
Gao, X. Y.; Yuan, Z. S.; Han, J. L. ; Wen, Z. L. ; Shan, S. S., Three New Spiral Galaxies with Active Nuclei Producing Double Radio Lobes, {\em Research in Astronomy and Astrophysics}, {\bf 2023}, {\em 23}, 3, 035005, 8

\bibitem[]{Yuan}
Yuan, Z. S.; Gao, X. Y.; Wen, Z. L. ; Han, J. L., Four Late-type Galaxies with Double Radio Lobes and Properties of Such Galaxies, {\em Research in Astronomy and Astrophysics}, {\bf 2024}, {\em 24}, 4, 045007, 7  

\bibitem[]{ODea2009}
O'Dea, C. P. ; Daly, R. A.; Kharb, P. ; Freeman, K. A. ; Baum, S. A., Physical properties of very powerful FRII radio galaxies,  {\em A\&A}, {\bf 2009}, {\em  494}, 2, 471-488

\bibitem[]{Gopal-Krishna}
Gopal-Krishna ; Wiita, Paul J., Asymmetries in Powerful Extragalactic Radio Sources, Asian Journal of Physics, 2004 ( https://arxiv.org/abs/astro-ph/0409761 )

\bibitem[]{GopalKrishnaWiita}
Gopal-Krishna; Wiita, Paul J., On the Origin of Correlated Radio-optical Asymmetries in Double Radio Sources, {\bf 1996}, {\em ApJ}, 467, 191

\bibitem[]{Freeland}
E. Freeland and E. Wilcots, Intergalactic Gas in Groups of Galaxies: Implications for Dwarf Spheroidal Formation and the Missing Baryons Problem, {\em ApJ}, {\bf 2011}, 738, 2, 145, 9 

 
\bibitem[]{Ho124Kantharia}
Kantharia, N. G. ; Ananthakrishnan, S. ; Nityananda, R. ; Hota, A, GMRT observations of the group Holmberg 124: Evolution by tidal forces and ram pressure? {\em A\&A,}, {\bf 2005}, {\em 435}, 2, 483-496

\bibitem[]{SaikiaDDRG}
Saikia, D.J. \& Jamrozy, M.,  Recurrent activity in Active Galactic Nuclei, {\em Bull. Astr. Soc. India}, {\bf 2009}, {\em 37}, 63-89

\bibitem[]{sagan5}
Dabhade, P., Chavan, K., Saikia, D. J., Oei, M. S., \& Röttgering, H. J., Search and analysis of giant radio galaxies with associated nuclei (SAGAN)-V. Study of giant double-double radio galaxies from LoTSS DR2, {\em A\&A}, {\bf 2025},  {\em 696}, A97.


\bibitem[]{HummelSaikia}
Hummel, E. \& Saikia, D. J., The anomalous radio features in NGC 4388 and NGC 4438, 
{\em A\&A}, {\bf 1991}, {\em  249}, 43-56.

\bibitem[]{NGC4569}
Chyzy, K. T.  ;  Soida, M.  ;  Bomans, D. J. ; Balkowski, Ch. ; Beck, R. ; Urbanik, M. Large-scale magnetized outflows from the Virgo Cluster spiral NGC4569, {\em A\&A}, {\bf 2006}, {\em 447}, 2, 465-47

\bibitem[]{MukherjeeFeedback}
Mukherjee, Dipanjan ; Bicknell, Geoffrey V. ; Wagner, Alexander Y. ; Sutherland, Ralph S.  ; Silk, Joseph, Relativistic jet feedback - III. Feedback on gas discs, {\em MNRAS}, {\bf 2018}, {\em 479}, 4, 5544-5566


\end{thebibliography}

\begin{thebibliography}{999}
\bibitem[Aranceta-Bartrina(1999a)]{ref-journal}
Aranceta-Bartrina, Javier. 1999a. Title of the cited article. \textit{Journal Title} 6: 100--10.
\bibitem[Aranceta-Bartrina(1999b)]{ref-book1}
Aranceta-Bartrina, Javier. 1999b. Title of the chapter. In \textit{Book Title}, 2nd ed. Edited by Editor 1 and Editor 2. Publication place: Publisher, vol. 3, pp. 54–96.
\bibitem[Baranwal and Munteanu {[1921]}(1955)]{ref-book2}
Baranwal, Ajay K., and Costea Munteanu. 1955. \textit{Book Title}. Publication place: Publisher, pp. 154--96. First published 1921 (op-tional).
\bibitem[Berry and Smith(1999)]{ref-thesis}
Berry, Evan, and Amy M. Smith. 1999. Title of Thesis. Level of Thesis, Degree-Granting University, City, Country. Identifi-cation information (if available).
\bibitem[Cojocaru et al.(1999)]{ref-unpublish}
Cojocaru, Ludmila, Dragos Constatin Sanda, and Eun Kyeong Yun. 1999. Title of Unpublished Work. \textit{Journal Title}, phrase indicating stage of publication.
\bibitem[Driver et al.(2000)]{ref-proceeding}
Driver, John P., Steffen Rohrs, and Sean Meighoo. 2000. Title of Presentation. In \textit{Title of the Collected Work} (if available). Paper presented at Name of the Conference, Location of Conference, Date of Conference.
\bibitem[Harwood(2008)]{ref-url}
Harwood, John. 2008. Title of the cited article. Available online: URL (accessed on Day Month Year).
\end{thebibliography}


\isAPAandChicago{}{%

}

\isChicagoStyle{%

}{}

\isAPAStyle{%

}{}

%


\PublishersNote{}
\end{adjustwidth}
\end{document}